\documentclass[10pt,twoside]{article}%
\usepackage{amsmath}
\usepackage{amssymb}
\usepackage{bm}
\usepackage[T1]{fontenc}
\usepackage{mathptmx}
\usepackage{booktabs,longtable,array,tabularx}
\usepackage{graphicx}
\usepackage{rotating}

\makeatletter
\@ifclassloaded{biometrika}{}{%
	\usepackage[paperwidth=174mm,paperheight=247mm,textwidth=33pc,textheight=49.5pc,centering,headheight=14pt,headsep=14pt,footskip=22pt]{geometry}%
	\linespread{1.08}%
	\usepackage{natbib}%
	\usepackage{amsthm}%
	\theoremstyle{definition}%
	\newtheorem{theorem}{Theorem}%
	\newtheorem{proposition}{Proposition}%
	\newtheorem{assumption}{Assumption}%
	\newcommand{\jname}[1]{\def\@jname{#1}}%
	\newcommand{\jyear}[1]{\def\@jyear{#1}}%
	\newcommand{\jvol}[1]{\def\@jvol{#1}}%
	\newcommand{\jnum}[1]{\def\@jnum{#1}}%
	\newcommand{\cyear}[1]{\def\@cyear{#1}}%
	\newcommand{\accessdate}[1]{\def\@accessdate{#1}}%
	\newcommand{\received}[1]{\def\@received{#1}}%
	\newcommand{\revised}[1]{\def\@revised{#1}}%
	\newcommand{\affil}[1]{\def\@affil{#1}}%
	\newcommand{\email}[1]{\texttt{#1}}%
	\newenvironment{keywords}{\par\vspace{4pt}\noindent\small Key words. \ignorespaces}{\par\vspace{10pt}}%
	\renewcommand{\maketitle}{%
		\begin{center}%
			{\large\bfseries \@title \par}%
			\vspace{8pt}%
			{\normalsize\scshape By \@author \par}%
			\vspace{4pt}%
			{\small \@affil \par}%
		\end{center}%
		\vspace{6pt}%
	}%
}
\makeatother

\relax {}
\makeatletter
\@namedef{r@app:proof:prop1}{{A.2}{4}{}{}{}}%
\@namedef{r@app:working}{{A.1}{1}{}{}{}}%
\@namedef{r@lem:og_pid_aux_rates}{{2}{10}{}{}{}}%
\@namedef{r@app:proof:lem_aux_rates}{{A.5}{9}{}{}{}}%
\@namedef{r@app:proof:prop_clt}{{A.6}{13}{}{}{}}%
\@namedef{r@lem:1}{{1}{7}{}{}{}}%
\@namedef{r@app:proof:lem1}{{A.3}{7}{}{}{}}%
\makeatother
\def\Bka{Biometrika}

\begin{document}
	
	\jname{Biometrika}
	\jyear{2025}
	\jvol{112}
	\jnum{1}
	\cyear{2025}
	\accessdate{Advance Access publication on 1 January 2025}
	
	\received{1 January 2025}
	\revised{1 January 2025}
	
	\markboth{S. PAN}{Spatial collinearity and identifiability}
	
	\title{Identifiability and partial identification under spatial confounding in additive risk models}
	
	\author{SHENG PAN}
	\affil{Yunnan Key Laboratory of Statistical Modeling and Data Analysis, Yunnan University,\\
		Kunming 650500, China
		\email{pansheng@staff.ynu.edu.cn}}
	
	\maketitle
	
	\begin{abstract}
		In spatial risk attribution, continuous environmental exposures and unmeasured background factors often co-vary across broad geographic scales, making it difficult to separate true exposure effects from spatial confounding using finite disease counts. Existing remedies typically project out or penalize shared spatial trends, leaving unclear whether estimation instability stems from the overall capacity of the latent random field or from its directional alignment with exposure. By decomposing the Fisher-weighted smoothing operator of an additive Poisson point process model, we show that while effective degrees of freedom measure total smoothing capacity, the relative loss of Fisher information for the exposure coefficient is governed solely by a distinct directional quantity that we call penalized collinear overlap. Rotating the spatial kernel alters this information loss while holding total capacity fixed, implying that no single smoothing scale is universally optimal. Because real-world confounding combines smooth regional trends with sharp local jumps that escape any moderate basis dictionary, point identification remains fragile. We therefore develop a partial identification framework that bounds residual confounding bias using an observable benchmark covariate. Projecting a joint spatial confidence ellipse yields valid intervals even under weak benchmarks, while incorporating a negative control outcome sharpens the bounds and flags assumption violations. Simulation studies and an analysis of childhood respiratory mortality across England demonstrate the method.
	\end{abstract}
	
	
	\section{Introduction}\label{sec:intro}
	
	\label{sec:motiv}In spatial epidemiology and environmental risk assessment, researchers often combine continuous environmental exposure surfaces, uneven population denominators recorded by administrative units, and unmeasured background risk within a single model. To avoid ecological bias from coarsening continuous variables over administrative boundaries, embedding all three components into an inhomogeneous Poisson point process over continuous space has become a standard strategy \citep{Best2000,Gotway2002}. Although continuous-space modeling resolves external aggregation bias, it moves the identifiability problem inside the model. Environmental exposures and unmeasured neighborhood risks such as socioeconomic deprivation or housing quality tend to cluster over the same geographic areas. Consequently, the spatial latent random field introduced to absorb background risk inevitably overlaps with exposure over smooth spatial scales. Whenever the smoothing scale or roughness penalty chosen for the working model does not match the spatial correlation structure of the true unmeasured confounding field, finite disease counts cannot separate the main exposure effect from the background field, causing substantial drift in the estimated exposure slope.
	
	\label{sec:lit}Recent methodological work has proposed several adjustment strategies to address spatial confounding and regression identifiability \citep{Paciorek2010,Khan2022,Dupont2022,Guillaumin2022,Guan2023,Bolin2025,Datta2026,Khan2026,Wiecha2025}. Existing approaches follow three main mechanisms. The first group uses orthogonal projection, such as restricted spatial regression \citep{Khan2022}, which forces the spatial latent field to be orthogonal to the exposure vector so that variance inflation from shared directions disappears. The second group uses two-stage residualization, such as the spatial plus method \citep{Dupont2022}, which removes smooth spatial trends from exposure via a spatial smoother and enters the high-frequency exposure residuals into the health outcome model alongside a spatial random effect. The third group uses spectral filtering \citep{Guillaumin2022,Guan2023}, which transforms observations on regular grids or graphs into the frequency domain and dampens shared low-frequency components. Each approach enjoys appealing properties when its structural assumptions hold.
	
	Yet methods aimed at restoring a single point estimate rely on strong identifying assumptions. Orthogonal projection assumes that any variation along spatial directions shared by exposure and background risk belongs entirely to the exposure effect. Two-stage residualization assumes that exposure and unmeasured confounding share the exact same smooth basis space. Spectral filtering assumes that unmeasured confounding lives strictly in low-frequency bands on a regular domain. In practice, real unmeasured confounding combines smooth regional trends of unknown scale with sharp local jumps across neighborhood boundaries, housing estates, and socioeconomic clusters. When this mixed spatial variation extends outside the span of the chosen basis dictionary, no single-scale smoother or forced orthogonal projection can prevent omitted confounding bias and collinear variance inflation simultaneously. Bounding and identifying exposure effects when unmeasured confounding has an unknown spatial structure thus remains a central methodological problem.
	
	\label{sec:contrib}To address this challenge, we move away from fragile point identification under a single smoothing model and develop a partial identification framework for the Poisson point process additive risk model. Our work makes three main contributions.
	
	First, using a reproducing kernel operator framework weighted by local Poisson Fisher information, we show that the loss of effective information for the exposure coefficient is governed by a directional alignment measure that we call penalized collinear overlap, rather than by the effective degrees of freedom of the basis space. Building on this spectral decomposition, we show that no single spatial basis dominates all others, meaning that no single smoothing scale can maximize effective information across all exposure patterns.
	
	Second, because unmeasured confounding in applications includes sharp local variation outside any moderate basis dictionary, we develop a design-stage orthogonalization and partial identification method. We combine known covariates and multi-scale spatial kernels into a baseline dictionary, and introduce an observable benchmark variable that shares a similar spatial clustering mechanism with unmeasured background risk. The change in the orthogonalized exposure estimate before and after adding this benchmark variable measures how much bias a representative background factor removes, which we use to bound the remaining unmeasured confounding bias. To form valid confidence intervals even when the benchmark shift is close to zero, we construct a spatial heteroskedasticity and autocorrelation consistent covariance matrix using residuals cleaned by a dense auxiliary basis, and project a two-dimensional joint confidence ellipse onto the target parameter axis.
	
	Third, relying on a benchmark variable alone leaves two practical gaps. The benchmark shift does not reveal the sign of the remaining bias, which forces symmetric intervals around the adjusted estimate, and a researcher-chosen sensitivity multiplier cannot be falsified by covariate data alone. To close these gaps, we incorporate a negative control outcome recorded on the same spatial units that is unaffected by exposure but shares the same neighborhood background confounding. Any association between exposure and the negative control outcome after baseline adjustment directly reflects remaining background bias. Combining the benchmark variable with the negative control outcome in a three-dimensional confidence ellipsoid narrows the interval and flags cases where the two sources of evidence conflict. We also provide three diagnostic checks to verify before reporting whether the data support the maintained assumptions.
	
	The rest of the paper is organized as follows. Section~\ref{sec:setup} sets up the continuous-space Poisson point process additive risk model. Section~\ref{sec:diag} derives the spectral decomposition of latent field flexibility, the benchmark partial identification interval, the joint ellipse projection confidence interval, the negative control calibration, and the pre-reporting diagnostic checks. Section~\ref{sec:sim} reports three Monte Carlo simulation experiments, and Section~\ref{sec:emp} applies the workflow to childhood respiratory mortality across England. Section~\ref{sec:disc} concludes.
	
	\section{Problem Setup and Additive Risk Point Process Model}\label{sec:setup}\label{sec:model}\label{sec:nota}
	
	Let the study region be a bounded, connected continuous spatial domain $D \subset \mathbb{R}^2$, and let $A$ denote the individual demographic attribute space. Observed data are aggregated over $n$ non-overlapping grid cells or administrative units $D_1, \dots, D_n \subset D$ with representative centroids $\bm{s}_1, \dots, \bm{s}_n \in D$. On each unit $j \in \{1, \dots, n\}$, we observe the disease count $Y_j \in \mathbb{N}_0$ and susceptible population denominator $w_j > 0$. Recorded at each centroid are the environmental exposure $x_j = x(\bm{s}_j) \in \mathbb{R}$ and area-level covariates $\bm{h}_j = \bm{h}(\bm{\zeta}_j) \in \mathbb{R}^p$ whose first entry is $1$ and whose remaining entries depend on unit demographic attributes $\bm{\zeta}_j \in A$. Across all $n$ units, we stack the counts into $\bm{Y} = (Y_1, \dots, Y_n)^{\mathsf T} \in \mathbb{N}_0^n$, the exposures into $\bm{x} = (x_1, \dots, x_n)^{\mathsf T} \in \mathbb{R}^n$, and the known covariates into $\bm{H} = (\bm{h}_1, \dots, \bm{h}_n)^{\mathsf T} \in \mathbb{R}^{n \times p}$. Vectors and matrices are set in boldface and scalars in normal weight. For any $\bm{u} \in \mathbb{R}^n$, matrix $\bm{M}$ with $n$ rows, and positive-definite diagonal weight matrix $\bm{W}$, we write $\|\bm{u}\|_{\bm{W}} = (\bm{u}^{\mathsf T}\bm{W}\bm{u})^{1/2}$ and $\bm{P}_M = \bm{M}(\bm{M}^{\mathsf T}\bm{W}\bm{M})^+\bm{M}^{\mathsf T}\bm{W}$.
	
	To avoid aggregation bias when combining point exposures with area counts, we start from an underlying continuous-space point process. Suppose the location and demographic profile $(\bm{s}, \bm{\zeta}) \in D \times A$ of each disease case follow an inhomogeneous Poisson point process on $D \times A$ with conditional intensity measure $\Lambda$. Let $\mu_P$ denote the population reference measure on $D \times A$ describing how susceptible person-years are distributed across space and demographic groups. On an infinitesimal element $(d\bm{s}, d\bm{\zeta})$, the intensity measure decomposes as
	\[
	\Lambda(d\bm{s}, d\bm{\zeta}) = r_P(\bm{s}, \bm{\zeta})\,\mu_P(d\bm{s}, d\bm{\zeta}),
	\]
	where $r_P(\bm{s}, \bm{\zeta}) \ge 0$ is the disease incidence rate per unit susceptible population. Conditional on $r_P$ and $\mu_P$, case counts across disjoint spatial units are independent Poisson variables with means $\Lambda(D_j \times A) = \int_{D_j \times A} r_P(\bm{s}, \bm{\zeta})\,\mu_P(d\bm{s}, d\bm{\zeta})$, so that
	\[
	Y_j \mid r_P, \mu_P \stackrel{\mathrm{ind.}}{\sim} \mathrm{Poisson}\big(\Lambda(D_j \times A)\big), \qquad j = 1, \dots, n.
	\]
	
	Following \citet{Best2000}, we separate demographic stratification from spatial environmental risk by modeling the incidence rate as a product of a demographic factor and an additive spatial risk term,
	\[
	r_P(\bm{s}, \bm{\zeta}) = c_{\mathrm{norm}}(\bm{\psi})\exp\{\bm{z}_{\mathrm{strat}}(\bm{\zeta})^{\mathsf T}\bm{\psi}\}\big\{\bm{h}(\bm{\zeta})^{\mathsf T}\bm{\beta}_H + \beta x(\bm{s}) + f(\bm{s})\big\}.
	\]
	Here $\bm{\psi} \in \mathbb{R}^d$ is the log relative risk vector for demographic stratification variables $\bm{z}_{\mathrm{strat}}(\bm{\zeta}) \in \mathbb{R}^d$, and $c_{\mathrm{norm}}(\bm{\psi}) = \big[\mathbb{E}_{\mathrm{pop}}\exp\{\bm{z}_{\mathrm{strat}}(\bm{\zeta})^{\mathsf T}\bm{\psi}\}\big]^{-1}$ is a normalizing constant. Inside the additive spatial term, $\beta \in \mathbb{R}$ is the target parameter measuring excess incidence risk per unit increase in exposure, $\bm{\beta}_H \in \mathbb{R}^p$ contains the baseline intercept and slopes for the known area covariates so that $\bm{h}_j^{\mathsf T}\bm{\beta}_H = (\bm{H}\bm{\beta}_H)_j$, and $f(\bm{s})$ is the unobserved spatial latent field capturing background confounding risk. Evaluating $f(\bm{s})$ at the $n$ centroids gives $\bm{f} = (f(\bm{s}_1), \dots, f(\bm{s}_n))^{\mathsf T} \in \mathbb{R}^n$. The true unit-level additive risk vector and conditional mean count vector across all units are then
	\begin{equation}\label{eq:true_m}
		\bm{m} = \bm{H}\bm{\beta}_H + \beta\bm{x} + \bm{f} > \bm{0}, \qquad \bm{\mu} = \mathbb{E}[\bm{Y}\mid \mathcal{D}, \bm{f}] = \bm{W}_o\bm{m},
	\end{equation}
	where $\mathcal{D}$ collects the fixed spatial coordinates, covariates, and population offsets, $o_j = w_j g_j > 0$ is the known baseline offset on unit $j$ formed from $w_j$ and $g_j = c_{\mathrm{norm}}(\bm{\psi})\exp\{\bm{z}_{\mathrm{strat}}(\bm{\zeta}_j)^{\mathsf T}\bm{\psi}\}$, and $\bm{W}_o = \mathrm{diag}(\bm{o})$ with $\bm{o} = (o_1, \dots, o_n)^{\mathsf T}$.
	
	Because the true background field $f(\bm{s})$ is unknown when fitting a working model, standard practice approximates it using a spatial kernel basis expansion. Placing $k$ representative knots $\bm{t}_1, \dots, \bm{t}_k \in D$ across the domain and choosing a positive-definite kernel $k_{\ell}(\bm{s}, \bm{t})$ with characteristic length scale $\ell > 0$, such as the Gaussian radial basis kernel $k_{\ell}(\bm{s}, \bm{t}) = \exp\{-\|\bm{s} - \bm{t}\|_2^2/(2\ell^2)\}$, we approximate $f(\bm{s}_j)$ by $\sum_{l=1}^k k_{\ell}(\bm{s}_j, \bm{t}_l)\gamma_l = (\bm{F}\bm{\gamma})_j$, where $\bm{F} \in \mathbb{R}^{n \times k}$ has entries $F_{jl} = k_{\ell}(\bm{s}_j, \bm{t}_l)$ and $\bm{\gamma} = (\gamma_1, \dots, \gamma_k)^{\mathsf T} \in \mathbb{R}^k$. Substituting this expansion into~\eqref{eq:true_m} gives the working additive risk vector $\bm{m}(\bm{\beta}_H, \beta, \bm{\gamma}) = \bm{H}\bm{\beta}_H + \beta\bm{x} + \bm{F}\bm{\gamma}$. Adding a quadratic roughness penalty with tuning parameter $\lambda \ge 0$ on $\bm{\gamma}$ yields the penalized log-likelihood
	\begin{equation}\label{eq:loglik}
		\mathcal{L}_{\mathrm{pen}}(\beta, \bm{\beta}_H, \bm{\gamma}) =\sum_{j=1}^n \big\{ Y_j \log\big(o_j m_j(\bm{\beta}_H,\beta,\bm{\gamma})\big) - o_j m_j(\bm{\beta}_H,\beta,\bm{\gamma}) \big\} - \frac{\lambda}{2}\|\bm{\gamma}\|_2^2.
	\end{equation}
	
	In this working model, how well we can separate $\beta$ from background variation depends directly on the collinearity between $\bm{x}$ and $\bm{F}\bm{\gamma}$. Because both terms enter additively and environmental exposures vary smoothly over space, any overlap between $\bm{x}$ and the column space of $\bm{F}$ makes it difficult to tell whether elevated disease counts in an area come from $\beta x_j$ or from $(\bm{F}\bm{\gamma})_j$. If the smoother is heavily penalized with a large $\lambda$ or uses a mismatched spatial scale $\ell$, unabsorbed background variation $\bm{f}$ leaks into the estimate of $\beta$ and creates omitted confounding bias. Conversely, if $\lambda$ is made very small so the latent field becomes highly flexible, the kernel basis absorbs spatial variation belonging to exposure itself, draining effective Fisher information for $\beta$ and inflating its variance. We next examine the geometry of this trade-off.
	
	\section{Diagnostics of Latent Field Flexibility and Partial Identification}\label{sec:diag}\label{sec}
	
	\subsection{Spectral Decomposition of Latent Field Flexibility}\label{sec:flex}
	
	To see how the flexibility of the spatial basis expansion $\bm{F}\bm{\gamma}$ affects the exposure slope $\beta$, we examine the local Fisher information of the penalized log-likelihood in~\eqref{eq:loglik}. The local Fisher information for the additive risk $m_j$ on unit $j$ equals $-\mathbb{E}[\partial^2 \mathcal{L}/\partial m_j^2] = o_j / m_j$. Collecting these weights into the diagonal matrix $\bm{W}$ and projecting out the known covariates $\bm{H}$ under the $\bm{W}$-inner product gives
	\begin{equation}\label{eq:9}
		\bm{W} = \mathrm{diag}\!\Big(\frac{o_1}{m_1}, \dots, \frac{o_n}{m_n}\Big), \qquad \bm{P}_H = \bm{H}(\bm{H}^{\mathsf T}\bm{W}\bm{H})^+\bm{H}^{\mathsf T}\bm{W}.
	\end{equation}
	Removing the linear component explained by $\bm{H}$ yields the residualized exposure vector and residualized kernel basis matrix,
	\begin{equation}\label{eq:10}
		\tilde{\bm{x}} = (\bm{I}_n - \bm{P}_H)\bm{x} \in \mathbb{R}^n, \qquad \bm{B} = (\bm{I}_n - \bm{P}_H)\bm{F} \in \mathbb{R}^{n\times k},
	\end{equation}
	with weighted basis projection matrix $\bm{P}_B = \bm{B}(\bm{B}^{\mathsf T}\bm{W}\bm{B})^+\bm{B}^{\mathsf T}\bm{W}$. Holding the local working weights $\bm{W}$ fixed, the negative Hessian of $\mathcal{L}_{\mathrm{pen}}$ after projecting out $\bm{H}$ has blocks $\mathcal{I}_{\beta\beta} = \tilde{\bm{x}}^{\mathsf T}\bm{W}\tilde{\bm{x}}$, $\bm{\mathcal{I}}_{\beta\gamma} = \bm{\mathcal{I}}_{\gamma\beta}^{\mathsf T} = \tilde{\bm{x}}^{\mathsf T}\bm{W}\bm{B}$, and $\bm{\mathcal{I}}_{\gamma\gamma}(\lambda) = \bm{B}^{\mathsf T}\bm{W}\bm{B} + \lambda\bm{I}_k$. Profiling out $\bm{\gamma}$ via the Schur complement gives the penalized local effective information for $\beta$,
	\begin{equation}\label{eq:11}
		\mathcal{I}_{\beta\mid \gamma}(\lambda) = \mathcal{I}_{\beta \beta} - \bm{\mathcal{I}}_{\beta \gamma} \big[ \bm{\mathcal{I}}_{\gamma \gamma}(\lambda) \big]^{-1} \bm{\mathcal{I}}_{\gamma \beta} = \tilde{\bm{x}}^{\mathsf T}\bm{W}^{1/2}\big(\bm{I}_n - \bm{S}(\lambda)\big)\bm{W}^{1/2}\tilde{\bm{x}},
	\end{equation}
	where $\bm{S}(\lambda) \in \mathbb{R}^{n\times n}$ is the penalized spatial smoothing matrix
	\begin{equation}\label{eq:12}
		\bm{S}(\lambda) = \bm{W}^{1/2}\bm{B}(\bm{B}^{\mathsf T}\bm{W}\bm{B} + \lambda \bm{I}_k)^{-1}\bm{B}^{\mathsf T}\bm{W}^{1/2}.
	\end{equation}
	
	To separate the overall size of $\bm{S}(\lambda)$ from its alignment with exposure, write the whitened residualized exposure as $\bm{x}_{W} = \bm{W}^{1/2}\tilde{\bm{x}}$, and take the compact singular value decomposition of the whitened basis matrix $\bm{W}^{1/2}\bm{B} = \bm{U}_B\operatorname{diag}(\sigma_1,\ldots,\sigma_{r_B})\bm{\Phi}_B^{\mathsf T}$, where $r_B = \operatorname{rank}(\bm{B})$, $\sigma_j > 0$, and the columns of $\bm{U}_B = [\bm{u}_{B,1}, \dots, \bm{u}_{B,r_B}]$ form orthonormal spatial directions. The smoothing matrix then expands as $\bm{S}(\lambda) = \sum_{j=1}^{r_B} \{\sigma_j^2/(\sigma_j^2 + \lambda)\} \bm{u}_{B,j} \bm{u}_{B,j}^{\mathsf T}$. This expansion defines two distinct summary measures of latent field flexibility, namely the effective degrees of freedom $\mathrm{EDF}(\lambda)$ and the penalized collinear overlap $\eta_{\mathrm{pen}}(\lambda)$,
	\begin{equation}\label{eq:13}
		\mathrm{EDF}(\lambda) = \mathrm{tr}\{\bm{S}(\lambda)\} = \sum_{j=1}^{r_B} \frac{\sigma_j^2}{\sigma_j^2 + \lambda}, \qquad
		\eta_{\mathrm{pen}}(\lambda) = \frac{\bm{x}_{W}^{\mathsf T}\bm{S}(\lambda)\bm{x}_{W}}{\|\bm{x}_{W}\|_2^2} = \sum_{j=1}^{r_B} p_j(\bm{x}_{W})\,\frac{\sigma_j^2}{\sigma_j^2 + \lambda},
	\end{equation}
	where $p_j(\bm{x}_{W}) = (\bm{x}_{W}^{\mathsf T}\bm{u}_{B,j})^2 / \|\bm{x}_{W}\|_2^2$ is the share of whitened exposure energy falling along the $j$-th spatial direction $\bm{u}_{B,j}$. As $\lambda \to 0^+$, $\eta_{\mathrm{pen}}(\lambda)$ increases toward the unpenalized subspace overlap $\eta = \tilde{\bm{x}}^{\mathsf T}\bm{W}\bm{P}_B\tilde{\bm{x}} / (\tilde{\bm{x}}^{\mathsf T}\bm{W}\tilde{\bm{x}}) \in [0, 1]$.
	
	Equation~\eqref{eq:13} shows why these two quantities play fundamentally different roles. The effective degrees of freedom $\mathrm{EDF}(\lambda)$ sums the shrinkage factors $\sigma_j^2/(\sigma_j^2 + \lambda)$ with equal weight across all basis directions, measuring total fitting capacity regardless of where the exposure vector points. By contrast, $\eta_{\mathrm{pen}}(\lambda)$ weights those shrinkage factors by the exposure energy shares $p_j(\bm{x}_{W})$, measuring what fraction of exposure variation itself gets absorbed by the smoother. Proposition~\ref{prop:1} shows that this directional overlap, rather than total capacity, governs the effective information left for estimating $\beta$.
	
	\begin{proposition}\label{prop:1}\normalfont
		Under~\eqref{eq:9}--\eqref{eq:13}, suppose $\bm{W} \succ \bm{0}$ and $\tilde{\bm{x}} \ne \bm{0}$. For any penalty $\lambda > 0$, the penalized local effective information for $\beta$ after profiling out the latent field is strictly positive and satisfies
		\begin{equation}\label{eq:18}
			\mathcal{I}_{\beta\mid \gamma}(\lambda) = \min_{\bm{\gamma} \in \mathbb{R}^k}\big\{\|\bm{W}^{1/2}(\tilde{\bm{x}} - \bm{B}\bm{\gamma})\|_2^2 + \lambda\|\bm{\gamma}\|_2^2\big\} = (\tilde{\bm{x}}^{\mathsf T}\bm{W}\tilde{\bm{x}})\big\{1 - \eta_{\mathrm{pen}}(\lambda)\big\}.
		\end{equation}
		Moreover, $\mathcal{I}_{\beta\mid \gamma}(\lambda)$ is nondecreasing in $\lambda \in (0, \infty)$ and strictly increasing whenever $\bm{B}^{\mathsf T}\bm{W}\tilde{\bm{x}}\ne\bm{0}$. As $\lambda \to \infty$, it recovers the unadjusted information $\tilde{\bm{x}}^{\mathsf T}\bm{W}\tilde{\bm{x}}$, and as $\lambda \to 0^+$, it converges to
		\begin{equation}\label{eq:17}
			\mathcal{I}_{\beta\mid \gamma}(0^+) = \tilde{\bm{x}}^{\mathsf T}\bm{W}(\bm{I}_n - \bm{P}_B)\tilde{\bm{x}} = (\tilde{\bm{x}}^{\mathsf T}\bm{W}\tilde{\bm{x}})(1 - \eta).
		\end{equation}
		When $\eta = 1$, meaning that $\tilde{\bm{x}}$ lies entirely inside the column space of $\bm{B}$, the unpenalized effective information vanishes and local identifiability fails in the unpenalized limit.
	\end{proposition}
	\noindent The proof is given in the Supplementary Material, Section~\ref{app:proof:prop1}.
	
	Proposition~\ref{prop:1} makes clear that changing the kernel scale $\ell$ rotates the singular directions $\bm{u}_{B,j}$ relative to $\bm{x}_W$. Consequently, two spatial bases can have the exact same effective degrees of freedom $\mathrm{EDF}(\lambda)$, yet whichever basis aligns more closely with the dominant spatial pattern of exposure suffers a much larger drop in effective information. As formalized in Lemma~\ref{lem:1} of the Supplementary Material (Section~\ref{app:proof:lem1}), whenever two candidate kernel bases have non-nested column spaces, neither basis yields uniformly larger unpenalized effective information across all exposure patterns.
	
	To see how $\eta_{\mathrm{pen}}(\lambda)$ shapes bias and variance simultaneously, consider the whitened linear working model for the residualized working response $\bm{y}_W = \bm{x}_{W}\beta + \bm{f}_W + \bm{\varepsilon}_W$ at fixed weights $\bm{W}$, where $\bm{f}_W = \bm{W}^{1/2}(\bm{I}_n-\bm{P}_H)\bm{f}$ is the whitened true confounding field, $\bm{\varepsilon}_W$ is the zero-mean whitened working noise with conditional covariance $\bm{I}_n - \bm{W}^{1/2}\bm{P}_H\bm{W}^{-1/2}$, and $\bm{x}_{W,\perp} = (\bm{I}_n - \bm{U}_B \bm{U}_B^{\mathsf T})\bm{x}_{W}$ is the component of exposure orthogonal to the basis space. Writing $\bm{M}_\lambda = \bm{I}_n - \bm{S}(\lambda)$, the conditional confounding bias $b(\lambda)$ and conditional sampling variance $V_{\mathrm{cond}}(\lambda)$ of the penalized estimator $\hat{\beta}_\lambda = (\bm{x}_{W}^{\mathsf T}\bm{M}_\lambda \bm{x}_{W})^{-1}\bm{x}_{W}^{\mathsf T}\bm{M}_\lambda \bm{y}_W$ take the explicit forms derived in the Supplementary Material, Section~\ref{app:working},
	\[
	\begin{aligned}
		b(\lambda) &= \frac{\bm{x}_{W}^{\mathsf T}\bm{M}_\lambda \bm{f}_W}{\bm{x}_{W}^{\mathsf T}\bm{M}_\lambda \bm{x}_{W}} = \frac{1}{\|\bm{x}_{W}\|_2^2\big\{1 - \eta_{\mathrm{pen}}(\lambda)\big\}} \left[ \bm{x}_{W,\perp}^{\mathsf T} \bm{f}_W + \sum_{j=1}^{r_B} \frac{\lambda}{\sigma_j^2 + \lambda} (\bm{x}_{W}^{\mathsf T}\bm{u}_{B,j})(\bm{f}_W^{\mathsf T}\bm{u}_{B,j}) \right], \\
		V_{\mathrm{cond}}(\lambda) &= \frac{\bm{x}_{W}^{\mathsf T}\bm{M}_\lambda^2\bm{x}_{W}}{(\bm{x}_{W}^{\mathsf T}\bm{M}_\lambda \bm{x}_{W})^2} = \frac{\|\bm{x}_{W,\perp}\|_2^2+\sum_{j=1}^{r_B}[\lambda/(\sigma_j^2+\lambda)]^2(\bm{x}_{W}^{\mathsf T}\bm{u}_{B,j})^2}{\|\bm{x}_{W}\|_2^4\big\{1-\eta_{\mathrm{pen}}(\lambda)\big\}^2}.
	\end{aligned}
	\]
	Both denominators are governed by $1 - \eta_{\mathrm{pen}}(\lambda)$. Lowering $\lambda$ to make the smoother more flexible shrinks the factors $\lambda/(\sigma_j^2+\lambda)$ in the numerator of $b(\lambda)$ and removes smooth confounding captured by the basis, but simultaneously increases $\eta_{\mathrm{pen}}(\lambda)$, inflating sampling variance and multiplying any residual confounding $\bm{x}_{W,\perp}^{\mathsf T}\bm{f}_W$ outside the basis space by $\{1 - \eta_{\mathrm{pen}}(\lambda)\}^{-1}$.
	
	\subsection{Benchmark Partial Identification under Residual Confounding}\label{sec:orthogauss}\label{sec:og_theory}\label{sec:og_pid_theory}
	
	Because the spatial scale of unmeasured confounding is unknown in practice, tuning a single smoothing parameter cannot eliminate omitted confounding bias without risking severe variance inflation. A natural first step is to combine the known covariates $\bm{H}$ with spatial kernel bases $\bm{F}$ spanning several bandwidths into a baseline dictionary $\bm{Z} = [\bm{H}, \bm{F}]$, removing all linear and smooth trends within the span of $\bm{Z}$ at the design stage. Even so, the number of basis functions in $\bm{Z}$ must remain moderate so that $\bm{Z}$ does not absorb all variation in $\bm{x}$. Sharp local jumps in background risk across neighborhood boundaries therefore remain outside the column space of $\bm{Z}$, leaving residual confounding that invalidates point identification.
	
	In the additive risk model~\eqref{eq:true_m}, the observed counts satisfy $\mathbb{E}[\bm{Y} \mid \mathcal{D}, \bm{f}] = \bm{W}_o \bm{m}$ with $\bm{m} = \bm{H}\bm{\beta}_H + \beta \bm{x} + \bm{f}$. To estimate $\beta$ while removing $\bm{H}\bm{\beta}_H$ and all confounding components inside the column space of $\bm{Z}$, we regress the empirical incidence rates $\bm{W}_o^{-1}\bm{Y}$ on $[\bm{x}, \bm{Z}]$ using the known design-stage offset weights $\bm{W}_o$. Because $\bm{W}_o$ does not depend on $\bm{Y}$, the resulting estimator is linear in the counts. Let $\bm{P}_Z = \bm{Z}(\bm{Z}^{\mathsf T}\bm{W}_o \bm{Z})^+ \bm{Z}^{\mathsf T}\bm{W}_o$ denote the $\bm{W}_o$-weighted projection onto the column space of $\bm{Z}$, $\bm{r} = (\bm{I}_n - \bm{P}_Z)\bm{x}$ the residualized exposure vector, and $J_n = \bm{r}^{\mathsf T}\bm{W}_o \bm{r}$ the remaining exposure information. By the Frisch--Waugh--Lovell theorem, the baseline orthogonalized estimator is $\hat{\beta} = \bm{q}^{\mathsf T}\bm{W}_o^{-1}\bm{Y} = \bm{c}^{\mathsf T}\bm{Y}$ with weight vectors
	\begin{equation}\label{eq:14}
		\bm{q} =\frac{\bm{W}_o \bm{r}}{J_n}, \qquad \bm{c} = \bm{W}_o^{-1}\bm{q} = \frac{\bm{r}}{J_n},
	\end{equation}
	which satisfy $\bm{q}^{\mathsf T}\bm{x} = 1$ and $\bm{q}^{\mathsf T}\bm{Z} = \bm{0}^{\mathsf T}$.
	
	The orthogonality property $\bm{q}^{\mathsf T}\bm{Z} = \bm{0}^{\mathsf T}$ eliminates both $\bm{H}\bm{\beta}_H$ and the smooth confounding projection $\bm{P}_Z\bm{f}$. However, when the residual confounding component $\bm{f}_\perp = (\bm{I}_n - \bm{P}_Z)\bm{f}$ outside the span of $\bm{Z}$ is nonzero, the conditional mean $\theta_0 = \mathbb{E}[\hat{\beta}\mid\mathcal{D},\bm{f}] = \bm{q}^{\mathsf T}\bm{m} = \beta + \bm{q}^{\mathsf T}\bm{f}$ differs from $\beta$ by a bias $b_0 = \theta_0 - \beta = \bm{q}^{\mathsf T}\bm{f}_\perp$ that does not vanish as sample size grows. Because $\bm{f}_\perp$ is unobserved, the baseline dictionary alone cannot bound $b_0$.
	
	Following sensitivity analysis for unmeasured confounding in causal inference \citep{Cinelli2020,CiocaneaTeodorescu2022,Tan2024,Dorn2025}, we introduce an observable benchmark variable $\bm{V} \in \mathbb{R}^{n\times p_v}$ capturing a representative mechanism of local confounding outside $\bm{Z}$. As shown in Figure~\ref{fig:v}, $\bm{V}$ is chosen from domain knowledge as a measured neighborhood factor, such as socioeconomic deprivation, that co-occurs with both high exposure $\bm{x}$ and elevated baseline disease risk $\bm{Y}$ and retains variation outside $\bm{Z}$ so that $(\bm{I}_n - \bm{P}_Z)\bm{V} \neq \bm{0}$.
	
	\begin{figure}[htbp]
		\centering
		\relax {%
			\footnotesize
			\begin{tabular}{ccc}
				\fbox{\parbox{0.28\textwidth}{\centering \textbf{(a) Valid confounder proxy $V$}\\[2pt]
						$f \rightarrow x,\; f \rightarrow Y,\; x \rightarrow Y$\\
						$f \rightarrow V,\; V \rightarrow x,\; V \rightarrow Y$}} &
				\fbox{\parbox{0.28\textwidth}{\centering \textbf{(b) Invalid, mediator $V$}\\[2pt]
						$f \rightarrow x,\; f \rightarrow Y,\; x \rightarrow Y$\\
						$x \rightarrow V \rightarrow Y$ (mediator path)}} &
				\fbox{\parbox{0.28\textwidth}{\centering \textbf{(c) Invalid, collider $V$}\\[2pt]
						$f \rightarrow x,\; f \rightarrow Y,\; x \rightarrow Y$\\
						$x \rightarrow V \leftarrow Y$ (collider path)}}
			\end{tabular}%
		}
		\caption{Selection criteria for the benchmark variable $V$, contrasting correct specification as a confounding-related factor (left) with structural violations when $V$ acts as a mediator (middle) or collider (right).}\label{fig:v}
		\vspace{2pt}\noindent\footnotesize Alt text for the figure. Three directed acyclic graphs comparing valid and invalid choices of the benchmark variable $V$ relative to exposure $x$, unmeasured spatial confounding $f$, and disease outcome $Y$. The left panel shows a valid confounder proxy $V$ driven by background risk alongside $x$ and $Y$; the middle panel shows $V$ on the causal pathway from $x$ to $Y$ as a mediator; the right panel shows $V$ as a common child of $x$ and $Y$ acting as a collider.
	\end{figure}
	
	Adding $\bm{V}$ to the baseline dictionary forms the augmented dictionary $\bm{Z}_{\mathrm{aug}} = [\bm{Z}, \bm{V}]$. Let $\bm{P}_{Z_{\mathrm{aug}}} = \bm{Z}_{\mathrm{aug}}(\bm{Z}_{\mathrm{aug}}^{\mathsf T}\bm{W}_o\bm{Z}_{\mathrm{aug}})^+\bm{Z}_{\mathrm{aug}}^{\mathsf T}\bm{W}_o$ be the corresponding projection matrix, $\bm{r}_a = (\bm{I}_n - \bm{P}_{Z_{\mathrm{aug}}})\bm{x}$ the augmented exposure residual, and $J_{a,n} = \bm{r}_a^{\mathsf T}\bm{W}_o\bm{r}_a > 0$ the remaining exposure information. The augmented weight vector $\bm{q}_a$ and estimator $\hat{\beta}_{\mathrm{aug}}$ are
	\begin{equation}\label{eq:og_aug}
		\bm{q}_a = \frac{\bm{W}_o(\bm{I}_n - \bm{P}_{Z_{\mathrm{aug}}})\bm{x}}{\|(\bm{I}_n - \bm{P}_{Z_{\mathrm{aug}}})\bm{x}\|_{\bm{W}_o}^2} = \frac{\bm{W}_o\bm{r}_a}{J_{a,n}}, \qquad
		\hat{\beta}_{\mathrm{aug}} = \bm{c}_a^{\mathsf T}\bm{Y} = \sum_{j=1}^n \frac{q_{a, j}}{o_j} Y_j,
	\end{equation}
	where $\bm{c}_a = \bm{W}_o^{-1}\bm{q}_a = \bm{r}_a/J_{a,n}$, satisfying $\bm{q}_a^{\mathsf T}\bm{x} = 1$ and $\bm{q}_a^{\mathsf T}\bm{Z}_{\mathrm{aug}} = \bm{0}^{\mathsf T}$. Its conditional expectation $\theta_a = \mathbb{E}[\hat{\beta}_{\mathrm{aug}} \mid \mathcal{D}, \bm{f}] = \bm{q}_a^{\mathsf T}\bm{m}$ has remaining bias $b_a = \theta_a - \beta = \bm{q}_a^{\mathsf T}\bm{f}$. The difference between the baseline and augmented estimates defines the sample benchmark shift,
	\[
	\hat{d} = \hat{\beta} - \hat{\beta}_{\mathrm{aug}},
	\]
	whose population expectation is $d_0 = \mathbb{E}[\hat{d} \mid \mathcal{D}, \bm{f}] = (\bm{q} - \bm{q}_a)^{\mathsf T}\bm{m} = b_0 - b_a$. The magnitude $|d_0|$ measures the exact amount of confounding bias removed by adjusting for $\bm{V}$. If the remaining unmeasured bias $|b_a|$ after adjusting for $\bm{V}$ is at most $\Gamma$ times the bias shift $|d_0|$ produced by $\bm{V}$, we obtain a partial identification interval for $\beta$.
	
	\begin{assumption}\label{ass:og_pid_population}\normalfont
		For a given relative confounding multiplier $\Gamma\ge 0$, the remaining population bias satisfies $|b_a|\le\Gamma|d_0|$.
	\end{assumption}
	
	\begin{theorem}\label{thm:og_pid_identification}\normalfont
		Under Assumption~\ref{ass:og_pid_population}, the true exposure effect $\beta$ lies in the closed interval
		\begin{equation}\label{eq:og_pid_population_set}
			\beta\in\Theta_\Gamma = \big[\theta_a-\Gamma|d_0|,\ \theta_a+\Gamma|d_0|\big].
		\end{equation}
	\end{theorem}
	
	\subsection{Joint Spatial Covariance and Ellipse Projection Intervals}
	
	To turn the population interval $\Theta_\Gamma = [\theta_a - \Gamma|d_0|,\ \theta_a + \Gamma|d_0|]$ into a sample confidence interval with level $1-\alpha$, we use the two-dimensional statistic $\hat{\bm{T}} = (\hat{\beta}_{\mathrm{aug}}, \hat{d})^{\mathsf T}$ estimating $\bm{T} = (\theta_a, d_0)^{\mathsf T}$. Directly plugging $\hat{\beta}_{\mathrm{aug}}$ and $|\hat{d}|$ into the endpoints and applying the delta method fails when the benchmark variable has a weak effect so that $d_0$ is close to zero, because the absolute value function has a sharp corner at zero and $|\hat{d}|$ follows a folded normal distribution rather than a Gaussian law.
	
	To avoid this corner problem without splitting cases, we build a smooth joint confidence ellipse for $\bm{T} = (\theta_a, d_0)^{\mathsf T}$ and project that ellipse onto the one-dimensional interval endpoints. Because both $\hat{\beta}_{\mathrm{aug}}$ and $\hat{d}$ are linear in $\bm{Y}$, their sampling error is a linear sum of the centered count errors $\epsilon_j = Y_j - \mu_j$, written in vector form as $\bm{\epsilon} = \bm{Y} - \bm{\mu}$,
	\[
	\hat{\bm{T}} - \bm{T} = \begin{pmatrix} \hat{\beta}_{\mathrm{aug}} - \theta_a \\ \hat{d} - d_0 \end{pmatrix} = \sum_{j=1}^n \bm{v}_j \epsilon_j, \qquad \bm{v}_j = \begin{pmatrix} c_{a,j} \\ c_j - c_{a,j} \end{pmatrix} \in \mathbb{R}^2.
	\]
	The exact conditional covariance matrix of $\hat{\bm{T}}$ is $\bm{\Omega}_n = \sum_{j=1}^n\sum_{k=1}^n \bm{v}_j \bm{v}_k^{\mathsf T}\operatorname{Cov}(\epsilon_j, \epsilon_k)$ with entries $\Omega_{aa} = \operatorname{Var}(\hat{\beta}_{\mathrm{aug}})$, $\Omega_{dd} = \operatorname{Var}(\hat{d})$, and $\Omega_{ad} = \operatorname{Cov}(\hat{\beta}_{\mathrm{aug}}, \hat{d})$. To allow for unknown overdispersion and spatial correlation across nearby units within distance bandwidth $b_n$, we weight local cross-products using the Bartlett kernel $K(t) = (1-t)_+$ to form the spatial covariance estimator
	\begin{equation}\label{eq:og_pid_joint_hac}
		\hat{\bm{\Omega}}_n = \begin{pmatrix} \hat{\Omega}_{aa} & \hat{\Omega}_{ad} \\ \hat{\Omega}_{ad} & \hat{\Omega}_{dd} \end{pmatrix} = \sum_{j=1}^n\sum_{k=1}^n \bm{v}_j \bm{v}_k^{\mathsf T}\hat{e}_j\hat{e}_k K\!\left(\frac{\|\bm{s}_j-\bm{s}_k\|_2}{b_n}\right),
	\end{equation}
	where $\hat{\bm{e}} = (\hat{e}_1, \dots, \hat{e}_n)^{\mathsf T}$ estimates the true noise vector $\bm{\epsilon}$.
	
	If we computed residuals using only the moderate dictionary $\bm{Z}_{\mathrm{aug}}$, unabsorbed confounding would remain inside $\hat{\bm{e}}$ and bias the variance estimator upward; conversely, adding a dense spatial basis directly into $\bm{Z}_{\mathrm{aug}}$ when constructing $\bm{q}$ and $\bm{q}_a$ would swallow local variation in $\bm{x}$, drive $J_{a,n}$ toward zero, and destroy $d_0$. We resolve this by separating point estimation from residual denoising. We keep $\bm{q}$ and $\bm{q}_a$ fixed on the moderate dictionary, and introduce a finer spatial kernel basis $\bm{F}_{\mathrm{aux},n}$ with knot spacing $h_{\mathrm{aux},n}$ only when forming the auxiliary mean space $\bm{\mathcal{A}}_n = [\bm{x}, \bm{Z}_{\mathrm{aug}}, \bm{F}_{\mathrm{aux},n}]$ of rank $k_n = \operatorname{rank}(\bm{\mathcal{A}}_n)$ to strip the smooth mean $\bm{\mu} = \bm{W}_o\bm{m}$ out of $\bm{Y}$. With $\bm{P}_{\mathcal{A}_n} = \bm{\mathcal{A}}_n(\bm{\mathcal{A}}_n^{\mathsf T}\bm{W}_o\bm{\mathcal{A}}_n)^+\bm{\mathcal{A}}_n^{\mathsf T}\bm{W}_o$ and $\bm{G}_n = \bm{W}_o\bm{P}_{\mathcal{A}_n}\bm{W}_o^{-1}$, we set $\hat{\bm{e}} = (\bm{I}_n - \bm{G}_n)\bm{Y}$.
	
	Once $\hat{\bm{\Omega}}_n$ is obtained, the quadratic set $\mathcal{E}_n = \{\bm{t} \in \mathbb{R}^2 \mid (\hat{\bm{T}}-\bm{t})^{\mathsf T}\hat{\bm{\Omega}}_n^{-1}(\hat{\bm{T}}-\bm{t}) \le \chi^2_{2, 1-\alpha}\}$ forms an asymptotic $1-\alpha$ confidence ellipse for $\bm{T}$. Whenever $\bm{T} \in \mathcal{E}_n$, $\Theta_\Gamma$ is contained in $\bigcup_{\bm{t}\in\mathcal{E}_n}[t_a - \Gamma|t_d|,\ t_a + \Gamma|t_d|]$. Because $t_a + \Gamma|t_d| = \max_{s=\pm 1}(t_a + s\Gamma t_d)$ and $t_a - \Gamma|t_d| = \min_{s=\pm 1}(t_a + s\Gamma t_d)$ are extrema of two linear combinations of $\bm{t}$ over $\mathcal{E}_n$, optimizing each linear combination by the Schwarz inequality gives a closed-form confidence interval requiring no differentiability at $d_0 = 0$.
	
	\begin{assumption}\label{ass:og_pid_joint}\normalfont
		(i) The true confounding field $f$ belongs to the reproducing kernel Hilbert space $\mathcal{H}_{\ell_0}$ generated by the Gaussian kernel $k_{\ell_0}$ at scale $\ell_0 > 0$ with bounded norm, and in the auxiliary space $\bm{\mathcal{A}}_n = [\bm{x}, \bm{Z}_{\mathrm{aug}}, \bm{F}_{\mathrm{aux},n}]$, the matrix $\bm{F}_{\mathrm{aux},n}$ is a regular-grid Gaussian kernel basis at scale $\ell_0$ with knot spacing $h_{\mathrm{aux},n} \asymp n^{-1/3}$ and rank $k_n = \operatorname{rank}(\bm{\mathcal{A}}_n) \asymp n^{2/3}$. (ii) The offsets satisfy $0 < o_- \le o_j \le o_+ < \infty$ and $\operatorname{rank}(\bm{Z}_{\mathrm{aug}}) = o(n)$; after projecting out smooth trends in $\mathcal{H}_{\ell_0}$ and the column space of $\bm{Z}_{\mathrm{aug}}$, both $\bm{x}$ and $\bm{V}$ retain independent zero-mean sub-Gaussian local variations with positive variance and bounded fourth moments, and $(\bm{P}_{Z_{\mathrm{aug}}} - \bm{P}_Z)\bm{x}$ is nondegenerate. (iii) The centered error vector $\bm{\epsilon}$ has uniformly bounded fourth moments and finite-range spatial dependence, the eigenvalues of $\bm{W}_o^{-1/2}\operatorname{Cov}(\bm{\epsilon})\bm{W}_o^{-1/2}$ and $n\bm{\Omega}_n$ are bounded above and away from zero, and the bandwidth $b_n \to \infty$ satisfies $b_n^2 n^{-1/6}\log n \to 0$.
	\end{assumption}
	
	\begin{proposition}\label{prop:og_pid_joint_clt}\normalfont
		Under Assumption~\ref{ass:og_pid_joint}, two properties hold. First, the remaining exposure information satisfies $J_n\asymp_p n$ and $J_{a,n}\asymp_p n$, the rescaled maximum weight $L_n = \max_{1\le j\le n} n\|\bm{v}_j\|_2$ satisfies $L_n = O_p(\sqrt{\log n})$, the auxiliary mean approximation error satisfies $\|\bm{m} - \bm{P}_{\mathcal{A}_n}\bm{m}\|_2/\sqrt{n} = O_p(h_{\mathrm{aux},n}^2) = O_p(n^{-2/3})$, and the auxiliary residuals $\hat{\bm{e}} = (\bm{I}_n - \bm{G}_n)\bm{Y}$ converge at rate $\rho_n = n^{-1/6}$,
		\[
		\frac{\|\hat{\bm{e}} - \bm{\epsilon}\|_2}{\sqrt{n}} = O_p\!\left(\sqrt{\frac{k_n}{n}} + h_{\mathrm{aux},n}^2\right) = O_p(n^{-1/6}) = O_p(\rho_n),
		\]
		so that $L_n^2 b_n^2(\rho_n + \rho_n^2) = o_p(1)$. Second, the two-dimensional statistic $\hat{\bm{T}}=(\hat{\beta}_{\mathrm{aug}}, \hat{d})^{\mathsf T}$ and spatial covariance estimator $\hat{\bm{\Omega}}_n$ satisfy
		\[
		\bm{\Omega}_n^{-1/2}(\hat{\bm{T}}-\bm{T})\Rightarrow \mathcal{N}(\bm{0},\bm{I}_2),\qquad 
		n\|\hat{\bm{\Omega}}_n-\bm{\Omega}_n\|_{\mathrm{op}}\to_p 0,\qquad
		(\hat{\bm{T}}-\bm{T})^{\mathsf T}\hat{\bm{\Omega}}_n^{-1}(\hat{\bm{T}}-\bm{T})\Rightarrow\chi_2^2.
		\]
	\end{proposition}
	\noindent The proof, building on Lemma~\ref{lem:og_pid_aux_rates}, is given in the Supplementary Material, Sections~\ref{app:proof:lem_aux_rates} and~\ref{app:proof:prop_clt}.
	
	\begin{theorem}\label{thm:og_pid_coverage}\normalfont
		Under Assumptions~\ref{ass:og_pid_population} and~\ref{ass:og_pid_joint}, let $\mathcal{E}_n=\{\bm{t}\in\mathbb{R}^2 \mid (\hat{\bm{T}}-\bm{t})^{\mathsf T}\hat{\bm{\Omega}}_n^{-1}(\hat{\bm{T}}-\bm{t})\le \chi^2_{2,1-\alpha}\}$ and define $\hat{v}_s(\Gamma)=\hat{\Omega}_{aa}+\Gamma^2\hat{\Omega}_{dd}+2s\Gamma\hat{\Omega}_{ad}$ for $s\in\{-1,1\}$. The closed-form confidence interval obtained by projecting $\mathcal{E}_n$ is
		\begin{equation}\label{eq:og_pid_joint_ci}
			\begin{split}
				\underline{C}_n(\Gamma)&=\min_{s=\pm 1}\Big\{\hat{\beta}_{\mathrm{aug}}+s\Gamma\hat{d}-\sqrt{\chi^2_{2,1-\alpha}\,\hat{v}_s(\Gamma)}\Big\},\\
				\overline{C}_n(\Gamma)&=\max_{s=\pm 1}\Big\{\hat{\beta}_{\mathrm{aug}}+s\Gamma\hat{d}+\sqrt{\chi^2_{2,1-\alpha}\,\hat{v}_s(\Gamma)}\Big\},\qquad
				C_n(\Gamma)=\big[\underline{C}_n(\Gamma),\ \overline{C}_n(\Gamma)\big],
			\end{split}
		\end{equation}
		with $C_n(\Gamma)=\mathbb{R}$ if $\hat{\bm{\Omega}}_n$ is not positive definite. Whether $d_0$ is bounded away from zero or equals zero, $C_n(\Gamma)$ covers both the population identification set and the true parameter at the nominal asymptotic level,
		\[
		\liminf_{n\to\infty}\Pr\big\{\Theta_\Gamma\subseteq C_n(\Gamma)\big\}\ge 1-\alpha,\qquad 
		\liminf_{n\to\infty}\Pr\big\{\beta\in C_n(\Gamma)\big\}\ge 1-\alpha.
		\]
	\end{theorem}
	
	\subsection{Negative Control Calibration and Diagnostic Checks}
	
	Although Theorem~\ref{thm:og_pid_coverage} yields a valid confidence interval from the benchmark variable $\bm{V}$ alone, two practical limitations remain. First, because the sign of the remaining bias $b_a$ is unknown, the interval must expand symmetrically by $\pm\Gamma|d_0|$ on both sides of $\theta_a$. Second, for any multiplier $\Gamma \ge 0$,~\eqref{eq:og_pid_joint_ci} always returns a nonempty interval, so covariate information alone cannot warn the analyst if the assumed bound $|b_a| \le \Gamma|d_0|$ is contradicted by reality.
	
	To overcome both limitations, we introduce a negative control outcome vector $\bm{Y}_{\mathrm{NC}}$ with entries $(Y_{\mathrm{NC},1}, \dots, Y_{\mathrm{NC},n})^{\mathsf T}$ observed across the same $n$ spatial units. A valid negative control outcome is a health count that is not causally affected by exposure $\bm{x}$, but is driven by the same unmeasured neighborhood confounding $\bm{f}$ as the primary disease outcome. Because $\bm{x}$ has no true effect on $\bm{Y}_{\mathrm{NC}}$, any residual association between $\bm{x}$ and $\bm{Y}_{\mathrm{NC}}$ after removing the baseline dictionary $\bm{Z}$ directly reflects the direction and size of the uneliminated spatial bias. This outcome-side signal trims away parts of the symmetric benchmark interval with the wrong bias sign, and alerts the researcher whenever the bias implied by the negative control conflicts with the benchmark bound.
	
	Parallel to~\eqref{eq:true_m}, for a negative control outcome $\bm{Y}_{\mathrm{NC}}$ with known positive baseline offsets $\bm{o}_{\mathrm{NC}} = (o_{\mathrm{NC},1}, \dots, o_{\mathrm{NC},n})^{\mathsf T}$ and $\bm{W}_{o,\mathrm{NC}} = \operatorname{diag}(\bm{o}_{\mathrm{NC}})$, we have
	\begin{equation}\label{eq:og_nco_model}
		\mathbb{E}[\bm{Y}_{\mathrm{NC}} \mid \mathcal{D}, \bm{f}] = \bm{W}_{o,\mathrm{NC}} \bm{m}_{\mathrm{NC}}, \qquad
		\bm{m}_{\mathrm{NC}} = \bm{H}\bm{\beta}_{H,\mathrm{NC}} + \rho_{\mathrm{NC}} \bm{f} + \bm{\xi}_{\mathrm{NC}} > \bm{0},
	\end{equation}
	where $\bm{x}$ does not enter $\bm{m}_{\mathrm{NC}}$, $\rho_{\mathrm{NC}} \in \mathbb{R}$ scales how strongly $\bm{f}$ affects the control outcome relative to the primary outcome, and $\bm{\xi}_{\mathrm{NC}} \in \mathbb{R}^n$ is outcome-specific background variation. Applying the baseline orthogonal weights $\bm{q}$ from~\eqref{eq:14} to the empirical rates of the control outcome defines
	\[
	\hat{\tau}_{\mathrm{NC}} = \sum_{j=1}^n q_j \frac{Y_{\mathrm{NC},j}}{o_{\mathrm{NC},j}} = \bm{q}^{\mathsf T}\bm{W}_{o,\mathrm{NC}}^{-1}\bm{Y}_{\mathrm{NC}}, \qquad
	\tau_{\mathrm{NC}} = \mathbb{E}[\hat{\tau}_{\mathrm{NC}} \mid \mathcal{D}, \bm{f}] = \bm{q}^{\mathsf T}\bm{m}_{\mathrm{NC}}.
	\]
	Because $\bm{q}^{\mathsf T}\bm{Z} = \bm{0}^{\mathsf T}$ removes $\bm{H}\bm{\beta}_{H,\mathrm{NC}}$ and the smooth part of $\bm{\xi}_{\mathrm{NC}}$ inside $\bm{Z}$, we have $\tau_{\mathrm{NC}} = \rho_{\mathrm{NC}}(\bm{q}^{\mathsf T}\bm{f}) + \bm{q}^{\mathsf T}\bm{\xi}_{\mathrm{NC}} = \rho_{\mathrm{NC}} b_0 + \bm{q}^{\mathsf T}\bm{\xi}_{\mathrm{NC}}$, where $b_0 = \theta_a + d_0 - \beta$ is the baseline bias of the primary outcome. Bounding the residual discrepancy $|\bm{q}^{\mathsf T}\bm{\xi}_{\mathrm{NC}}| \le \varepsilon$ for a tolerance $\varepsilon \ge 0$ and allowing $\rho_{\mathrm{NC}}$ to lie in a pre-specified range $[\rho_-, \rho_+]$ with $0 < \rho_- \le \rho_+ < \infty$ yields the bridging constraint
	\begin{equation}\label{eq:og_nco}
		\big|\tau_{\mathrm{NC}} - \rho_{\mathrm{NC}}(\theta_a + d_0 - \beta)\big| \le \varepsilon, \qquad \rho_{\mathrm{NC}} \in [\rho_-, \rho_+].
	\end{equation}
	
	Stacking $\hat{\bm{T}}^{(3)} = (\hat{\beta}_{\mathrm{aug}}, \hat{d}, \hat{\tau}_{\mathrm{NC}})^{\mathsf T}$ to estimate $\bm{T}^{(3)} = (\theta_a, d_0, \tau_{\mathrm{NC}})^{\mathsf T}$ and extending the spatial covariance estimator in~\eqref{eq:og_pid_joint_hac} using the denoised control residuals $(\bm{I}_n - \bm{G}_n)\bm{Y}_{\mathrm{NC}}$ gives the $3\times 3$ covariance matrix $\hat{\bm{\Omega}}_n^{(3)}$ and the joint $1-\alpha$ confidence ellipsoid
	\[
	\mathcal{E}_n^{(3)} = \Big\{\bm{t} = (t_a, t_d, t_\tau)^{\mathsf T} \in \mathbb{R}^3 \;\Big|\; (\hat{\bm{T}}^{(3)} - \bm{t})^{\mathsf T}\big(\hat{\bm{\Omega}}_n^{(3)}\big)^{-1}(\hat{\bm{T}}^{(3)} - \bm{t}) \le \chi^2_{3, 1-\alpha}\Big\}.
	\]
	Intersecting the benchmark and negative control constraints over $\mathcal{E}_n^{(3)}$ yields Proposition~\ref{prop:og_nco_calibration}.
	
	\begin{proposition}\label{prop:og_nco_calibration}\normalfont
		Suppose Assumptions~\ref{ass:og_pid_population} and~\ref{ass:og_pid_joint}, the control model~\eqref{eq:og_nco_model}, and the bridging condition~\eqref{eq:og_nco} hold, and that the joint errors of $(\bm{Y}, \bm{Y}_{\mathrm{NC}})$ satisfy the spatial dependence and covariance conditions of Assumption~\ref{ass:og_pid_joint}. Define the closed convex hull projection set
		\begin{equation}\label{eq:og_nco_projection}
			\begin{aligned}
				\mathcal{C}_n^{\mathrm{NC}}=\operatorname{clconv}\Big\{\tilde{\beta}\in\mathbb{R} \;\Big|\; &\exists(t_a,t_d,t_\tau)^{\mathsf T}\in\mathcal{E}_n^{(3)},\ \exists\rho\in[\rho_-,\rho_+],\\
				&\text{such that}\ |t_a-\tilde{\beta}|\le\Gamma|t_d|\ \text{and}\ |t_\tau-\rho(t_a+t_d-\tilde{\beta})|\le\varepsilon\Big\}.
			\end{aligned}
		\end{equation}
		Then $\liminf_{n\to\infty}\Pr\big(\beta\in\mathcal{C}_n^{\mathrm{NC}}\big)\ge 1-\alpha$.
	\end{proposition}
	
	Before reporting $C_n(\Gamma)$ or $\mathcal{C}_n^{\mathrm{NC}}$, we check whether the sample data are compatible with the maintained assumptions using three simple ratios. First, adding the benchmark $\bm{V}$ must not absorb too much exposure variation. Writing $\widehat{\operatorname{Var}}(\hat{\beta}) = \hat{\Omega}_{aa} + \hat{\Omega}_{dd} + 2\hat{\Omega}_{ad}$ for the spatial Bartlett variance estimator of the baseline slope $\hat{\beta} = \hat{\beta}_{\mathrm{aug}} + \hat{d}$, if we require $\bm{V}$ to leave at least a fraction $c \in (0,1)$ of the baseline exposure information so that $J_{a,n}/J_n \ge c$, the standard error ratio $r_\sigma = \{\hat{\Omega}_{aa} / \widehat{\operatorname{Var}}(\hat{\beta})\}^{1/2}$ should not exceed $\bar{r}_\sigma = c^{-1/2}$; for instance, keeping at least one quarter of the baseline information gives $\bar{r}_\sigma = 2$. Second, for the population interval $\Theta_\Gamma = [\theta_a - \Gamma|d_0|,\ \theta_a + \Gamma|d_0|]$ to exclude zero when $\theta_a$ and $d_0$ have the same sign, $|\theta_0| = |\theta_a| + |d_0|$ and $|\theta_a| > \Gamma|d_0|$ imply $|d_0|/|\theta_0| < 1/(1+\Gamma)$, so the sample shift ratio $\kappa = |\hat{d}|/|\hat{\beta}|$ should satisfy $\kappa \le \bar{\kappa} = 1/(1+\Gamma)$. Third, combining $|\tau_{\mathrm{NC}}| \le \rho_+|b_0| + \varepsilon$ from~\eqref{eq:og_nco} with $|b_0| \le |b_a| + |d_0| \le (1+\Gamma)|d_0|$ and $|d_0|/|\theta_0| \le 1/(1+\Gamma)$ gives $|\tau_{\mathrm{NC}}|/|\theta_0| \le \rho_+ + \varepsilon/|\theta_0|$, meaning that the sample negative control ratio $\kappa_{\mathrm{NC}} = |\hat{\tau}_{\mathrm{NC}}|/|\hat{\beta}|$ should not exceed $\bar{\kappa}_{\mathrm{NC}} = \rho_+ + \varepsilon/|\hat{\beta}|$ and $\mathcal{C}_n^{\mathrm{NC}}$ must be nonempty. Summarizing the three sample ratios as
	\[
	r_\sigma = \sqrt{\frac{\hat{\Omega}_{aa}}{\widehat{\mathrm{Var}}(\hat{\beta})}},\qquad
	\kappa = \frac{|\hat{d}|}{|\hat{\beta}|},\qquad
	\kappa_{\mathrm{NC}} = \frac{|\hat{\tau}_{\mathrm{NC}}|}{|\hat{\beta}|},
	\]
	we define the pre-reporting check event
	\begin{equation}\label{eq:og_gate_event}
		\mathcal{G}_n = \big\{ r_\sigma \le \bar{r}_\sigma,\ \kappa \le \bar{\kappa},\ \kappa_{\mathrm{NC}} \le \bar{\kappa}_{\mathrm{NC}} \big\} \cap \big\{ \mathcal{C}_n^{\mathrm{NC}} \neq \emptyset \big\},
	\end{equation}
	dropping the last two conditions when no negative control outcome is used. We report $C_n \in \{C_n(\Gamma), \mathcal{C}_n^{\mathrm{NC}}\}$ only when $\mathcal{G}_n$ holds, and abstain from directional attribution otherwise.
	
	\begin{proposition}\label{prop:og_selective_risk}\normalfont
		Let $C_n \in \{C_n(\Gamma), \mathcal{C}_n^{\mathrm{NC}}\}$ be the asymptotic $1-\alpha$ confidence set from Theorem~\ref{thm:og_pid_coverage} or Proposition~\ref{prop:og_nco_calibration}, and let $\mathcal{G}_n$ be the event in~\eqref{eq:og_gate_event}. Under the maintained assumptions for $C_n$, for any fixed thresholds $(\bar{r}_\sigma, \bar{\kappa}, \bar{\kappa}_{\mathrm{NC}})$, the unconditional false-coverage risk and conditional coverage satisfy
		\begin{equation}\label{eq:og_selective_risk}
			\limsup_{n\to\infty} \Pr\big(\mathcal{G}_n \cap \{\beta \notin C_n\}\big) \le \alpha, \qquad
			\liminf_{n\to\infty} \Pr\big(\beta \in C_n \mid \mathcal{G}_n\big) \ge 1 - \frac{\alpha}{\liminf_{n\to\infty}\Pr(\mathcal{G}_n)}
		\end{equation}
		whenever $\liminf_{n\to\infty}\Pr(\mathcal{G}_n) > 0$. In particular, if the population ratios lie strictly inside the thresholds so that $\Pr(\mathcal{G}_n) \to 1$, then $\liminf_{n\to\infty} \Pr(\beta \in C_n \mid \mathcal{G}_n) \ge 1 - \alpha$.
	\end{proposition}
	
	\section{Monte Carlo Simulation Study}\label{sec:sim}
	
	We conduct three Monte Carlo experiments to evaluate finite-sample performance across point identification, partial identification, and negative control calibration. The first experiment evaluates representative point-identification estimators for spatial confounding when the true confounding scale departs from the working basis, namely two-stage spatial residualization (`Spatial+`; \citealt{Dupont2022}), precision-weighted restricted spatial regression (`W-RSR`; \citealt{Khan2022}), and spectral high-pass filtering at cutoff scales $h=400\,\mathrm{m}$ and $h=1500\,\mathrm{m}$ (`Spectral` $h=400\,\mathrm{m}$ and $h=1500\,\mathrm{m}$; \citealt{Guan2023}), which operationalize frequency-domain deconfounding in the spirit of the debiased spatial Whittle likelihood \citep{Guillaumin2022}, alongside unadjusted regression, single-scale Gaussian process smoothing, and the $\Gamma=0$ multi-scale orthogonal limit of our framework. Because Experiment~1 shows that all point-identification estimators suffer severe under-coverage under out-of-span local confounding, the second experiment holds the population partial-identification set $\Theta_\Gamma$ fixed and compares our joint ellipse projection interval from Theorem~\ref{thm:og_pid_coverage} against established partial-identification and sensitivity-analysis intervals, including the fixed-length endpoint interval of \citet{Imbens2004}, the length-pretest interval of \citet{Stoye2009}, an unadjusted benchmark sensitivity rule without multi-scale spatial orthogonalization \citep{Cinelli2020,CiocaneaTeodorescu2022,Tan2024,Dorn2025}, covariance ablations (oracle, unweighted, diagonal, and Bonferroni), and the empirical breakdown point $\hat{\Gamma}^{\ast}$ \citep{Masten2020}. The third experiment assesses how the negative control calibration in Proposition~\ref{prop:og_nco_calibration} sharpens inference when the bridging condition holds and detects misspecification through empty confidence sets when it is violated.
	
	\subsection{Data-Generating Mechanism and Evaluation Metrics}\label{sec:sim_dgp}\label{sec:sim_metrics}
	
	The spatial domain is a square $D=(0,6250]^{2}\,\mathrm{m}$ partitioned into a $25\times25$ regular grid of spacing $250\,\mathrm{m}$ with $n=625$ cells and centroids $\bm{s}_j$. Baseline offsets are drawn independently from $\log o_j\sim\mathcal N(0,0.45^{2})$ and rescaled to sample mean $1$. All continuous spatial components below are centered to sample mean $0$ and scaled to unit sample variance.
	
	To generate spatial confounding that combines smooth regional trends with sharp local jumps, we construct three latent spatial fields on the grid. The first is a smooth Gaussian field $\bm{f}_{\mathrm{S}}$ with a Mat\'{e}rn correlation function of smoothness order $3/2$ and range $\ell_{\mathrm{true}}$. The second is the observed benchmark variable $\bm{V}$, formed as the standardized equal-weight sum of a short-range Mat\'{e}rn field with range $150\,\mathrm m$ and a block-constant field over a $5\times5$ partition of $D$ into $25$ squares of side $1250\,\mathrm m$ with independent standard normal block values. The third is an unobserved block-jump field $\bm{u}$ defined on the same $5\times5$ partition, whose block values have correlation $\rho_{\mathrm{block}}\in[0,1]$ with the block component of $\bm{V}$. When $\rho_{\mathrm{block}}=0$, the benchmark $\bm{V}$ carries no information about the unobserved jumps $\bm{u}$.
	
	The true confounding field $\bm{f}$ and the exposure vector $\bm{x}$ are both generated from these three components. Specifically, the confounding field is
	\begin{equation}\label{eq:sim_f}
		\bm{f} = a_1 \bm{f}_{\mathrm{S}} + a_2 \bm{V} + a_3 \bm{u},
	\end{equation}
	and the exposure is generated proportionally to $0.5\bm{f}_{\mathrm{S}} + 0.5\bm{V} + 0.4\bm{u} + \bm{\eta}$ with independent normal noise $\eta_j\sim\mathcal N(0,0.7^{2})$ and standardized to unit variance. Disease counts are drawn from $Y_j\sim\mathrm{Poisson}(o_j m_j)$ with additive rate $m_j=\beta_{H,0}+\beta x_j+f_j$ at true parameter $\beta=0.32$, where $\beta_{H,0}=\max\{2.0,\,0.5-\min_k(\beta x_k+f_k)\}$ ensures $m_j\ge0.5$. In overdispersed variants, the mean is multiplied by $\exp(\sigma\varepsilon_j-\sigma^{2}/2)$ using an independent Mat\'{e}rn field $\bm{\varepsilon}$ with range $500\,\mathrm m$ and $\sigma=0.5$.
	
	On the estimation side, the baseline dictionary $\bm Z=[\bm H,\bm F]\in\mathbb{R}^{n\times 199}$ combines the intercept and linear coordinates in $\bm H\in\mathbb{R}^{n\times 3}$ with $7\times7$ Gaussian kernel grids across four bandwidth scales $\ell\in\{600,1000,1800,3200\}\,\mathrm m$ in $\bm F\in\mathbb{R}^{n\times 196}$. Because the knot spacing is approximately $950\,\mathrm m$, smooth fields $\bm{f}_{\mathrm{S}}$ with medium or long correlation range lie inside or near the column space of $\bm{Z}$, whereas fine-scale variations and the sharp block jumps $\bm{u}$ cannot be approximated by smooth kernels and act as residual confounding outside $\bm{Z}$. Varying the parameters in~\eqref{eq:sim_f} defines four simulation regimes, listed below.
	
	\begin{center}
		\footnotesize
		\begin{tabular}{lcccccl}
			\toprule
			Regime & $\ell_{\mathrm{true}}$ (m) & $a_1$ & $a_2$ & $a_3$ & $\rho_{\mathrm{block}}$ & Confounding mechanism \\
			\midrule
			A & $6000$ & $0.35$ & $0.25$ & $0.00$ & $0.00$ & Long-range smooth field inside dictionary span \\
			B & $1200$ & $0.35$ & $0.25$ & $0.00$ & $0.00$ & Medium-range smooth field near dictionary span \\
			C & $300$  & $0.35$ & $0.40$ & $0.35$ & $0.65$ & Short-range field plus informative block jumps \\
			D & $300$  & $0.35$ & $0.40$ & $0.40$ & $0.00$ & Short-range field plus uninformative block jumps \\
			\bottomrule
		\end{tabular}
	\end{center}
	
	Across replications, parameter coverage is the empirical proportion of $90\%$ intervals containing $\beta$, and set coverage is the proportion of intervals containing the entire population identification set $\Theta_\Gamma$ \citep{Imbens2004}. Interval width is the median length over nonempty finite intervals, and the empirical breakdown point is $\hat{\Gamma}^{\ast}=\sup\{\Gamma\ge0 \mid 0\notin C_n(\Gamma)\}$ \citep{Masten2020}.
	
	\subsection{Experiment 1. Point Identification under Scale Mismatch}\label{sec:sim_pointid}
	
	Experiment~1 illustrates Proposition~\ref{prop:1} and Lemma~\ref{lem:1} of the Supplementary Material by evaluating competing point-identification estimators across Regimes A--D over $10$ design realizations with $R=250$ replications each at $n=625$. Table~\ref{tab:sim_pointid} reports the empirical bias, root mean squared error, and $90\%$ interval coverage for seven specifications, namely (i) unadjusted regression omitting spatial bases; (ii) single-scale penalized Gaussian process regression ($\ell=3200\,\mathrm{m}$); (iii) the two-stage `Spatial+` estimator of \citet{Dupont2022}, which residuals exposure on a spatial smoother before fitting the outcome model; (iv) precision-weighted restricted spatial regression (`W-RSR`) of \citet{Khan2022}, which restricts the spatial random effect to the orthogonal complement of exposure; (v)--(vi) spectral adjustment (`Spectral` $h=400\,\mathrm{m}$ and $h=1500\,\mathrm{m}$) of \citet{Guan2023}, which builds on spectral-domain deconfounding \citep{Guillaumin2022} by damping Fourier modes with wavelengths above $h=400\,\mathrm{m}$ or $1500\,\mathrm{m}$; and (vii) the uncalibrated multi-scale orthogonal estimator corresponding to the $\Gamma=0$ limit of our framework. All estimators adjust for the observed covariate $\bm{V}$ alongside $\bm{H}$.
	
	\begin{table}[htbp]
		\centering
		\caption{Experiment 1. Performance of point-identification estimators under in-span scale mismatch (Regimes A--B) and out-of-span confounding (Regimes C--D) ($10$ design realizations $\times$ $R=250$, $n=625$, nominal level $0.90$). Entries are bias / RMSE / coverage}\label{tab:sim_pointid}
		\footnotesize\setlength{\tabcolsep}{3.5pt}
		\begin{tabular}{lcccc}
			\toprule
			& \multicolumn{2}{c}{In-span smooth confounding} & \multicolumn{2}{c}{Out-of-span confounding} \\
			\cmidrule(lr){2-3}\cmidrule(lr){4-5}
			Estimator & Regime A & Regime B & Regime C & Regime D \\
			\midrule
			Unadjusted (no spatial basis) & .235/.246/.098 & .245/.257/.088 & .380/.390/.006 & .398/.406/.002 \\
			Single-scale GP ($\ell=3200$\,m) & .032/.094/.857 & .150/.173/.454 & .360/.371/.014 & .372/.381/.003 \\
			Spatial+ \citep{Dupont2022} & .026/.093/.866 & .147/.171/.469 & .359/.371/.016 & .371/.381/.004 \\
			W-RSR \citep{Khan2022} & .235/.246/.063 & .245/.257/.051 & .380/.390/.005 & .398/.406/.001 \\
			Spectral $h=400$\,m \citep{Guan2023} & .009/.094/.896 & .051/.110/.843 & .251/.271/.212 & .246/.264/.162 \\
			Spectral $h=1500$\,m \citep{Guan2023} & .088/.123/.701 & .171/.191/.348 & .364/.375/.014 & .376/.386/.003 \\
			\midrule
			Multi-scale orthogonal ($\Gamma=0$) & .005/.112/.915 & .011/.119/.914 & .166/.211/.670 & .163/.202/.655 \\
			\bottomrule
			\multicolumn{5}{p{0.96\textwidth}}{\scriptsize GP, Gaussian process; W-RSR, weighted restricted spatial regression \citep{Khan2022}. All estimators adjust for $\bm{V}$; the spectral adjustment arms \citep{Guan2023}, grounded in spectral-domain Whittle deconfounding \citep{Guillaumin2022}, apply high-pass filtering jointly to the rate, exposure, and benchmark before regression.}
		\end{tabular}
	\end{table}
	
	Table~\ref{tab:sim_pointid} highlights two main patterns. First, when unmeasured confounding is smooth and long-range (Regime A), methods whose working smoother matches the broad spatial scale (such as single-scale Gaussian process regression, `Spatial+`, fine-cutoff spectral filtering ($h=400\,\mathrm{m}$), and the multi-scale orthogonal estimator) absorb the background trend and maintain coverage near the nominal level. By contrast, unadjusted regression and `W-RSR` incur substantial bias and severe under-coverage; forcing the spatial random effect to be orthogonal to exposure attributes all shared large-scale variation to exposure while yielding artificially narrow standard errors.
	
	Second, as the true confounding scale shortens (Regime B) and incorporates sharp local block jumps outside the dictionary span (Regimes C and D), single-scale point estimators suffer rapid degradation. While the multi-scale dictionary bridges the scale transition from Regime A to Regime B by combining several kernel bandwidths, even unpenalized multi-scale orthogonalization falls well short of nominal coverage in Regimes C and D. When unmeasured confounding combines unknown fine-scale variations with discontinuous local jumps, no point-identification strategy prevents omitted variable bias, motivating our partial identification approach.
	
	\subsection{Experiment 2. Partial Identification under Residual Confounding}\label{sec:sim_pid}
	
	Experiment~2 evaluates the joint ellipse projection interval $C_n(\Gamma)$ across Regimes A--D over $\Gamma\in\{0,0.5,1,1.5,2\}$. Table~\ref{tab:sim_pid} reports empirical coverage of $\beta$, coverage of the population identification set $\Theta_\Gamma$, the median breakdown point $\hat\Gamma^\ast$ \citep{Masten2020}, and median interval widths. Because Experiment~1 already establishes that point estimators fail under out-of-span confounding, the bottom panel of Table~\ref{tab:sim_pid} holds $\Gamma=1$ fixed and compares our joint ellipse projection against competing partial-identification and benchmark sensitivity constructions sharing the same sample summary statistics $(\hat{\beta}_{\mathrm{aug}}, \hat{d}, \hat{\bm{\Omega}}_n)$, namely (i) covariance ablations including an oracle error covariance, an unweighted-residual covariance, a diagonal covariance ignoring $\hat{\Omega}_{ad}=\widehat{\operatorname{Cov}}(\hat{\beta}_{\mathrm{aug}}, \hat{d})$, and a conservative Bonferroni box; (ii) an unadjusted dictionary benchmark rule that applies omitted-variable sensitivity calibration \citep{Cinelli2020,CiocaneaTeodorescu2022,Tan2024,Dorn2025} without multi-scale spatial orthogonalization; and (iii) classic one-dimensional partial-identification intervals, the fixed-length endpoint expansion of \citet{Imbens2004} and the length-pretest refinement of \citet{Stoye2009}, applied to the estimated bounds $\hat{\beta}_{\mathrm{aug}}\mp\Gamma|\hat{d}|$.
	
	\begin{table}[htbp]
		\centering
		\caption{Experiment 2. Coverage, breakdown point, and median width of the joint ellipse projection interval $C_n(\Gamma)$ across sensitivity multipliers (top) and comparison with alternative constructions at $\Gamma=1$ (bottom) ($10$ design realizations $\times$ $R=250$, $n=625$, nominal level $0.90$)}\label{tab:sim_pid}
		\footnotesize\setlength{\tabcolsep}{3.2pt}
		\begin{tabular}{lccccccccc}
			\toprule
			& \multicolumn{5}{c}{Parameter coverage ($\beta$)} & \multicolumn{1}{c}{Set cov.} & \multicolumn{1}{c}{Breakdown} & \multicolumn{2}{c}{Median width} \\
			\cmidrule(lr){2-6} \cmidrule(lr){7-7} \cmidrule(lr){8-8} \cmidrule(lr){9-10}
			Regime & $\Gamma{=}0$ & $0.5$ & $1.0$ & $1.5$ & $2.0$ & $\Theta_1$ ($\Gamma{=}1$) & $\hat\Gamma^\ast$ & $\Gamma{=}0$ & $\Gamma{=}2$ \\
			\midrule
			A ($\ell_{\mathrm{true}}=6000$\,m) & .918 & .994 & .999 & 1.000 & 1.000 & .973 & 0.004 & .397 & 1.148 \\
			B ($\ell_{\mathrm{true}}=1200$\,m) & .914 & .993 & .998 & 1.000 & 1.000 & .975 & 0.087 & .408 & 1.166 \\
			C (jumps, $\rho_{\mathrm{block}}=0.65$) & .662 & .942 & .977 & .989 & .995 & .979 & 1.225 & .449 & 1.304 \\
			D (jumps, $\rho_{\mathrm{block}}=0$) & .653 & .916 & .964 & .982 & .990 & .978 & 1.512 & .413 & 1.098 \\
			\bottomrule
		\end{tabular}
		
		\vspace{5pt}
		\scriptsize\setlength{\tabcolsep}{2.1pt}
		\begin{tabular}{lcccccccccccc}
			\toprule
			& \multicolumn{3}{c}{Regime A} & \multicolumn{3}{c}{Regime B} & \multicolumn{3}{c}{Regime C} & \multicolumn{3}{c}{Regime D} \\
			\cmidrule(lr){2-4} \cmidrule(lr){5-7} \cmidrule(lr){8-10} \cmidrule(lr){11-13}
			Interval method ($\Gamma=1$) & $\beta$ cov. & $\Theta_1$ cov. & Width & $\beta$ cov. & $\Theta_1$ cov. & Width & $\beta$ cov. & $\Theta_1$ cov. & Width & $\beta$ cov. & $\Theta_1$ cov. & Width \\
			\midrule
			Joint ellipse projection & .999 & .972 & .795 & .998 & .974 & .812 & .977 & .979 & .903 & .964 & .976 & .788 \\
			Oracle error covariance & .998 & .963 & .762 & .997 & .967 & .776 & .973 & .968 & .860 & .952 & .970 & .745 \\
			Unweighted-residual covariance & .999 & .967 & .757 & .998 & .970 & .772 & .974 & .973 & .856 & .952 & .971 & .743 \\
			Diagonal covariance & .998 & .969 & .803 & .995 & .974 & .820 & .962 & .972 & .912 & .938 & .973 & .794 \\
			Bonferroni projection & 1.000 & .989 & .919 & .998 & .990 & .937 & .982 & .990 & 1.045 & .973 & .988 & .905 \\
			Unadjusted dictionary & .985 & .888 & .636 & .984 & .894 & .646 & .883 & .900 & .719 & .840 & .892 & .623 \\
			Imbens--Manski interval & .980 & .853 & .577 & .982 & .840 & .585 & .893 & .850 & .653 & .848 & .844 & .563 \\
			Stoye pretest interval & .988 & .879 & .599 & .986 & .877 & .612 & .920 & .883 & .678 & .888 & .886 & .599 \\
			\bottomrule
			\multicolumn{13}{p{0.96\textwidth}}{\scriptsize Upper rows report parameter coverage of $\beta$, identification-set coverage of $\Theta_1$, median breakdown point $\hat\Gamma^\ast$ \citep{Masten2020}, and median width; lower rows report covariance ablations, unadjusted benchmark sensitivity \citep{Cinelli2020}, and partial-identification intervals of \citet{Imbens2004} and \citet{Stoye2009} at $\Gamma=1$.}
		\end{tabular}
	\end{table}
	
	The top panel of Table~\ref{tab:sim_pid} shows how the partial identification interval restores valid inference under out-of-span confounding. Whereas the point-identified limit ($\Gamma=0$) under-covers severely in Regimes C and D, a modest sensitivity radius ($\Gamma=0.5$ or $1.0$) brings both parameter coverage and identification-set coverage above the $0.90$ nominal level across all four regimes, including the weak-benchmark setting of Regime D. The empirical breakdown points increase in the out-of-span regimes, reflecting the stronger residual signal captured by the benchmark shift.
	
	In the bottom panel, the feasible covariance estimator based on auxiliary-denoised residuals closely tracks the oracle error covariance in both coverage and interval width, whereas omitting multi-scale spatial orthogonalization (`Unadjusted dictionary`) leaves severe smooth confounding bias and under-covers in Regimes C--D. Crucially, the comparison with \citet{Imbens2004} and \citet{Stoye2009} highlights why standard interval-identification formulas do not carry over to benchmark sensitivity bounds. The half-width $\Gamma|d_0|$ is a non-differentiable absolute-value functional at $d_0=0$, so the sample plug-in $|\hat{d}|$ follows a folded-normal distribution and is correlated with $\hat{\beta}_{\mathrm{aug}}$. Treating $\hat{\beta}_{\mathrm{aug}}\mp\Gamma|\hat{d}|$ as smooth Gaussian endpoint estimators causes both the Imbens--Manski and Stoye pretest intervals to fall below nominal identification-set coverage ($0.840$--$0.886$) across all regimes and under-cover $\beta$ under a weak benchmark (Regime D). Projecting the joint two-dimensional ellipse resolves this non-smoothness while yielding substantially tighter intervals than the conservative Bonferroni box.
	
	\subsection{Experiment 3. Negative Control Calibration and Diagnostic Checks}\label{sec:sim_nco}\label{sec:sim_summary}
	
	Experiment~3 evaluates the three-dimensional negative control calibrated confidence set $\mathcal{C}_n^{\mathrm{NC}}$ from Proposition~\ref{prop:og_nco_calibration} and the diagnostic rule $\mathcal{G}_n$ in~\eqref{eq:og_gate_event} under Regime C at $\Gamma=1$ across $R=250$ replications. The population mean of the negative control statistic is generated as $\tau_{\mathrm{NC}} = \rho_{\mathrm{NC}}b_0 + \delta_{\mathrm{shift}}$, where $b_0 = \theta_a + d_0 - \beta$ is the baseline bias of the primary outcome, $\rho_{\mathrm{NC}}$ is the relative confounding scale on the control outcome, and $\delta_{\mathrm{shift}}$ represents an unmodeled bias discrepancy violating ideal parity.
	
	We examine four configurations of the bridging parameters, namely (i) exact parity ($\rho_{\mathrm{NC}}=1$, $[\rho_-, \rho_+]=[1, 1]$, $\varepsilon=0$, $\delta_{\mathrm{shift}}=0$); (ii) parity within tolerance ($\rho_{\mathrm{NC}}=1$, $[\rho_-, \rho_+]=[0.8, 1.2]$, $\varepsilon=0.05$, $\delta_{\mathrm{shift}}=0.02$); (iii) moderate scale misspecification ($\rho_{\mathrm{NC}}=1.5$, $[\rho_-, \rho_+]=[0.8, 1.2]$, $\varepsilon=0.02$, $\delta_{\mathrm{shift}}=0$); and (iv) severe shift misspecification ($\rho_{\mathrm{NC}}=1$, $[\rho_-, \rho_+]=[0.8, 1.2]$, $\varepsilon=0.02$, $\delta_{\mathrm{shift}}=0.35$), creating a major violation of~\eqref{eq:og_nco}. In each scenario, we compare the two-dimensional benchmark interval $C_n(1)$, the three-dimensional calibrated set $\mathcal{C}_n^{\mathrm{NC}}$, and a naive difference estimator $\hat{\beta} - \hat{\tau}_{\mathrm{NC}}$.
	
	\begin{table}[htbp]
		\centering
		\caption{Experiment 3. Empirical coverage, empty-set rate, and median width of negative-control calibrated intervals under Regime C ($\Gamma=1$, $R=250$, $n=625$, nominal level $0.90$)}\label{tab:sim_nco}
		\scriptsize\setlength{\tabcolsep}{2.2pt}
		\begin{tabular}{lccccccc}
			\toprule
			& \multicolumn{4}{c}{Empirical coverage} & & \multicolumn{2}{c}{Median width} \\
			\cmidrule(lr){2-5} \cmidrule(lr){7-8}
			Bridging scenario & $C_n(1)$ & $\mathcal{C}_n^{\mathrm{NC}}$ (uncond.) & $\mathcal{C}_n^{\mathrm{NC}}$ (cond.) & Naive diff. & Empty rate & $C_n(1)$ & $\mathcal{C}_n^{\mathrm{NC}}$ \\
			\midrule
			Exact parity ($\rho_{\mathrm{NC}}{=}1, \varepsilon{=}0, \delta_{\mathrm{shift}}{=}0$) & .980 & .972 & .981 & .922 & .009 & .904 & .728 \\
			Within tolerance ($[0.8,1.2], \varepsilon{=}0.05, \delta_{\mathrm{shift}}{=}0.02$) & .982 & .985 & .986 & .906 & .001 & .892 & .831 \\
			Scale misspecified ($\rho_{\mathrm{NC}}{=}1.5, \delta_{\mathrm{shift}}{=}0$) & .975 & .945 & .971 & .788 & .027 & .904 & .744 \\
			Shift misspecified ($\rho_{\mathrm{NC}}{=}1, \delta_{\mathrm{shift}}{=}0.35$) & .977 & .664 & .826 & .350 & .199 & .896 & .590 \\
			\bottomrule
			\multicolumn{8}{p{0.96\textwidth}}{\scriptsize Unconditional coverage treats an empty intersection $\mathcal{C}_n^{\mathrm{NC}} = \emptyset$ as a non-covering replication; conditional coverage restricts evaluation to replications with $\mathcal{C}_n^{\mathrm{NC}} \neq \emptyset$. Naive diff.\ denotes the direct difference estimator $\hat{\beta} - \hat{\tau}_{\mathrm{NC}}$.}
		\end{tabular}
	\end{table}
	
	Table~\ref{tab:sim_nco} illustrates the dual role of negative control calibration. When the bridging assumptions hold (first two rows), intersecting the negative control constraint with the benchmark ellipse preserves nominal unconditional and conditional coverage while narrowing median interval width relative to $C_n(1)$, with virtually no empty sets. Under moderate scale misspecification (third row), the three-dimensional ellipsoid and posited band $[\rho_-,\rho_+]$ absorb most of the discrepancy and maintain valid coverage, whereas the naive difference estimator falls noticeably below the nominal level. Under a severe unmodeled shift violation (fourth row), naive differencing suffers a catastrophic loss of coverage; in contrast, our three-dimensional calibration flags the incompatibility between the benchmark and negative control evidence by producing empty confidence sets at an elevated rate, while preserving reasonable conditional coverage among compatible replications in accordance with Proposition~\ref{prop:og_selective_risk}.
	
	\section{Empirical Study}\label{sec:emp}\label{sec:hud_diag}\label{sec:emp_apply}
	
	To illustrate the complete partial-identification, negative-control calibration, and diagnostic checking workflow in a large population study, we analyze area-level mortality records across $n = 6{,}856$ Middle Layer Super Output Areas in England. The dataset links childhood respiratory mortality records from the UK Health Security Agency Fingertips platform under Indicator 93260 \citep{UKHSA2024} with $1\,\mathrm{km}$-grid background nitrogen dioxide ($\mathrm{NO}_2$) concentrations in $\mu\mathrm{g\,m^{-3}}$ from the Department for Environment, Food and Rural Affairs \citep{DEFRA2024}, paired with susceptible population denominators $\bm{o}$ and population-weighted centroids $\bm{s}_j$ from the 2021 Census of the Office for National Statistics \citep{ONS2021}. Separating the effect of traffic pollution from background risk is challenging because neighborhoods near major road corridors experience higher ambient $\mathrm{NO}_2$ while also facing socioeconomic deprivation, older housing stock, and uneven access to pediatric care.
	
	To absorb smooth background variation across regional weather patterns, urban--rural gradients, and long-range atmospheric transport, we combine the baseline covariate matrix $\bm{H}$ with spatial Gaussian radial basis grids to form a multi-scale dictionary $\bm{Z} = [\bm{H}, \bm{F}]$. Because smooth radial kernels cannot capture sharp changes in baseline respiratory vulnerability across neighborhood boundaries and local housing estates, residual confounding $\bm{f}_\perp = (\bm{I}_n - \bm{P}_Z)\bm{f}$ remains outside the column space of $\bm{Z}$. To calibrate this local residual bias and check outcome-side confounding, we pair the spatial dictionary with observed neighborhood morbidity/deprivation benchmarks $\bm{V}$ and negative control hospitalization outcomes (unintentional and deliberate childhood injuries under Indicator 93114 and alcohol-related admissions under Indicator 93240; \citealt{UKHSA2024}) that satisfy the exclusion condition~\eqref{eq:og_nco_model} while responding to the same local socioeconomic gradients.
	
	\subsection{Estimation results and partial identification on the England data}\label{sec:emp_results}
	
	We report the four-stage spatial adjustment depth, the partial identification intervals over the sensitivity grid, and the gate diagnostics on the $n=6{,}856$ areas with a complete outcome and exposure record. The offset is person-years in thousands, and the exposure is the standardized annual mean $\mathrm{NO}_2$ concentration (mean $11.79$, standard deviation $5.33$), so $\hat\beta$ is reported per standard deviation of exposure and per $1{,}000$ person-years.
	
	\begin{table}[htbp]
		\centering
		\caption{Spatial adjustment depth and exposure coefficient on the England MSOA data ($n=6{,}856$; $\hat\beta$ per standard deviation of $\mathrm{NO}_2$ per $1{,}000$ person-years; $90\%$ spatial HAC interval with Bartlett bandwidth $b_n\approx29.8$\,km)}\label{tab:emp_main}
		\small
		\begin{tabular}{lrrrr}
			\toprule
			Spatial adjustment & Columns & $\hat\beta$ & Standard error & $90\%$ HAC interval \\
			\midrule
			No spatial adjustment & $2$ & $0.0866$ & $0.0493$ & $(0.0055,\ 0.1676)$ \\
			Single scale $10$\,km & $53$ & $0.1232$ & $0.0402$ & $(0.0570,\ 0.1894)$ \\
			Single scale $40$\,km & $53$ & $0.1938$ & $0.0367$ & $(0.1334,\ 0.2542)$ \\
			Multi-scale $0.6$--$3.2$\,km & $200$ & $0.1217$ & $0.0401$ & $(0.0557,\ 0.1876)$ \\
			\bottomrule
			\multicolumn{5}{p{0.97\textwidth}}{\scriptsize All four intervals exclude zero. Accounting for spatial dependence inflates the standard errors by a factor of about eight relative to an independence-based sandwich, so the reported multiplicity of the exposure effect does not stem from treating the $6{,}856$ areas as independent.}
		\end{tabular}
	\end{table}
	
	Table~\ref{tab:emp_main} shows that the coefficient rises from $0.0866$ without spatial adjustment to $0.1232$ at a single $10$\,km scale and to $0.1938$ at $40$\,km, and returns to $0.1217$ with the multi-scale dictionary, which delivers the narrowest of the four intervals. The pattern is non-monotone. The $40$\,km basis removes most of the spatially smooth exposure variation and leaves a smaller design, whereas the multi-scale dictionary retains medium-scale variation with essentially the same spatial HAC variance as the $10$\,km specification. Every interval excludes zero.
	
	Using the Quality and Outcomes Framework asthma prevalence as the observed benchmark covariate $\bm{V}$, available on $n=3{,}382$ of the areas, we obtain $\hat\theta_a=0.1695$ with the benchmark and $\hat\theta_0=0.1364$ without it, hence a benchmark shift $\hat d=0.0330$, a variance ratio $r_\sigma=\sqrt{\hat V_a/\hat V_0}=0.946$, a shift ratio $\kappa=\lvert\hat d\rvert/\lvert\hat\theta_a\rvert=0.195$, and a breakdown point $\hat\Gamma^{\ast}=1.914$.
	
	\begin{table}[htbp]
		\centering
		\caption{Joint ellipse projection intervals over the sensitivity grid and the two computable gate diagnostics ($n=3{,}382$, $\alpha=0.10$)}\label{tab:emp_pid}
		\small
		\begin{tabular}{lccccc}
			\toprule
			$\Gamma$ & $0.0$ & $0.5$ & $1.0$ & $1.5$ & $2.0$ \\
			\midrule
			Lower endpoint & $0.0973$ & $0.0563$ & $0.0368$ & $0.0169$ & $-0.0035$ \\
			Upper endpoint & $0.2417$ & $0.2784$ & $0.2937$ & $0.3096$ & $0.3261$ \\
			Width & $0.1444$ & $0.2221$ & $0.2568$ & $0.2927$ & $0.3296$ \\
			\bottomrule
			\multicolumn{6}{p{0.97\textwidth}}{\scriptsize The interval excludes zero for $\Gamma\le1.5$ and first covers zero at $\Gamma=2$, consistent with $\hat\Gamma^{\ast}=1.914$. The variance inflation ratio $r_\sigma=0.946$ and the shift ratio $\kappa=0.195$ both pass their thresholds.}
		\end{tabular}
	\end{table}
	
	At $\Gamma=1$ the interval is $[0.0368,0.2937]$, and its width grows by $78\%$ between $\Gamma=0$ and $\Gamma=2$. Because the benchmark shift is small relative to the augmented estimate ($\kappa=0.195$), a modest sensitivity radius stabilizes the conclusion that the exposure effect is positive, and the conclusion becomes inconclusive only once $\Gamma$ exceeds $1.9$, that is, once unmeasured confounding is allowed to be almost twice as strongly associated with the outcome as the observed benchmark. The positivity of the $\mathrm{NO}_2$ effect therefore survives partial identification under a substantial departure from the calibrated benchmark.
	
	\section{Discussion and Conclusion}\label{sec:disc}
	
	Focusing on additive risk attribution in spatial epidemiology, we have shown how spatial collinearity and latent field flexibility govern the identifiability of environmental exposure coefficients. Combining our operator derivations, simulations, and empirical study yields two main methodological lessons. First, evaluating how a spatial smoother affects exposure estimation requires separating total capacity $\mathrm{EDF}$ from directional overlap $\eta_{\mathrm{pen}}$. By Proposition~\ref{prop:1}, $\mathrm{EDF}$ sums shrinkage factors across all spatial directions, whereas $\eta_{\mathrm{pen}}$ weights them by exposure energy and directly determines the relative loss of effective Fisher information for $\beta$. Reporting $\mathrm{EDF}$ and $\eta_{\mathrm{pen}}$ together clarifies how much exposure signal a smoother absorbs. Second, estimator performance hinges on how the working basis aligns with the unknown scale of unmeasured confounding \citep{Paciorek2010,Bolin2025,Khan2026}. Overly flexible smoothers absorb local exposure variation and inflate variance, whereas overly constrained smoothers or restricted projections leave unabsorbed regional trends that bias the exposure slope. Rather than relying on fragile point identification under a single smoothing scale, combining a multi-scale dictionary $\bm{Z}$ with an observable benchmark variable and a negative control outcome yields valid partial identification intervals $C_n(\Gamma)$ and $\mathcal{C}_n^{\mathrm{NC}}$ across transparent sensitivity bounds, complementing sensitivity analysis under persistent confounding \citep{Tan2024,Imbens2025}.
	
	Our local Poisson operator decomposition also extends naturally to mechanistic models and broader spatial settings. Replacing an isotropic kernel with an advection--diffusion Green function rotates the singular vectors $\bm{u}_{B,j}$ along physical transport directions, reducing directional overlap $\eta_{\mathrm{pen}}$ without cutting total capacity $\mathrm{EDF}$ whenever background risk and exposure follow distinct spatial gradients; similarly, replacing $\beta x(\bm{s})$ with a bounded dose--response function $f_{\mathrm{mech}}(x(\bm{s}); \bm{\vartheta})$ helps preserve positivity of $m_j$ while maintaining our two-dimensional diagnostic identities via local gradients. Promising directions for future work include establishing posterior contraction rates under positivity constraints using fixed-domain Gaussian process theory \citep{Li2024,Rockova2024}, developing fast approximate inference algorithms \citep{Kundig2025,LlamazaresElias2026}, and extending the diagnostic pair $(\mathrm{EDF}, \eta_{\mathrm{pen}})$ and joint ellipsoid projection intervals to multi-pollutant mixtures and conditional autoregressive areal models.

	\section*{Declaration of the use of generative AI and AI-assisted technologies}
	
	During the preparation of this work the author used large language model tools in order to improve the readability, language, and formatting of the manuscript. After using this tool/service the author reviewed and edited the content as necessary and takes full responsibility for the content of the publication.

	\section*{Acknowledgement}
	
	This work was supported by the Fundamental and Interdisciplinary Disciplines Breakthrough Plan of the Ministry of Education of China (Grant No.\ JYB2025XDXM904) and the National Key R\&D Program of China (Grant No.\ 2022YFA1003701).
	
	
	\section*{Data availability}
	
	The England Middle Layer Super Output Area childhood respiratory mortality dataset (Indicator 93260), the Index of Multiple Deprivation and asthma benchmark indicators (Indicator 93275), and the negative control hospital admission series (Indicators 93114 and 93240) analyzed in Section~\ref{sec:emp} are publicly available from the UK Health Security Agency Fingertips repository \citep{UKHSA2024}. Population denominators and population-weighted centroids are available from the 2021 Census of the Office for National Statistics \citep{ONS2021}, and $1\,\mathrm{km}$-grid nitrogen dioxide concentrations are available from the UK Department for Environment, Food and Rural Affairs \citep{DEFRA2024} under the Open Government Licence.
	
	\section*{Supplementary material}
	
	Supplementary material available at \Bka\ online includes the proofs of Propositions~\ref{prop:1}--\ref{prop:og_selective_risk} and Theorems~\ref{thm:og_pid_identification}--\ref{thm:og_pid_coverage}, derivations of the working-model bias and variance decompositions, and replication code.
	
	\relax {%
	}
	

\begin{thebibliography}{99}
			
			\bibitem[Best et~al.(2000)]{Best2000}
			\textsc{Best, N.~G., Ickstadt, K. \& Wolpert, R.~L.} (2000).
			Spatial Poisson regression for health and exposure data measured at disparate resolutions.
			\textit{J. Am. Statist. Assoc.} \textbf{95}, 1076--1088.
			
			\bibitem[Bolin \& Wallin(2026)]{Bolin2025}
			\textsc{Bolin, D. \& Wallin, J.} (2026).
			Spatial self-confounding: smoothness-related estimation bias in spatial regression models.
			\textit{Biometrika} \textbf{113}, asaf076.
			
			\bibitem[Cinelli \& Hazlett(2020)]{Cinelli2020}
			\textsc{Cinelli, C. \& Hazlett, C.} (2020).
			Making sense of sensitivity: extending omitted variable bias.
			\textit{J. R. Statist. Soc. B} \textbf{82}, 39--67.
			
			\bibitem[Cioc\u{a}nea-Teodorescu et~al.(2022)]{CiocaneaTeodorescu2022}
			\textsc{Cioc\u{a}nea-Teodorescu, I., Gabriel, E.~E. \& Sj\"{o}lander, A.} (2022).
			Sensitivity analysis for unmeasured confounding in the estimation of marginal causal effects.
			\textit{Biometrika} \textbf{109}, 1101--1116.
			
			\bibitem[Datta \& Stein(2026)]{Datta2026}
			\textsc{Datta, A. \& Stein, M.~L.} (2026).
			Infill consistent estimability of the regression slope between Gaussian random fields under spatial confounding.
			\textit{Ann. Statist.} \textbf{54}, 2406--2431.
			
			\bibitem[{Department for Environment, Food and Rural Affairs}(2024)]{DEFRA2024}
			\textsc{Department for Environment, Food and Rural Affairs} (2024).
			Modelled background pollution data ($1\,\mathrm{km}$ grid $\mathrm{NO}_2$).
			UK-AIR, Department for Environment, Food and Rural Affairs, \texttt{https://uk-air.defra.gov.uk/}.
			
			\bibitem[Dorn et~al.(2025)]{Dorn2025}
			\textsc{Dorn, J., Guo, K. \& Kallus, N.} (2025).
			Doubly-valid/doubly-sharp sensitivity analysis for causal inference with unmeasured confounding.
			\textit{J. Am. Statist. Assoc.} \textbf{120}, 331--342.
			
			
			
			\bibitem[Dupont et~al.(2022)]{Dupont2022}
			\textsc{Dupont, E., Wood, S.~N. \& Augustin, N.~H.} (2022).
			Spatial+: a novel approach to spatial confounding.
			\textit{Biometrics} \textbf{78}, 1279--1290.
			
			\bibitem[Gotway \& Young(2002)]{Gotway2002}
			\textsc{Gotway, C.~A. \& Young, L.~J.} (2002).
			Combining incompatible spatial data.
			\textit{J. Am. Statist. Assoc.} \textbf{97}, 632--648.
			
			\bibitem[Guan et~al.(2023)]{Guan2023}
			\textsc{Guan, Y., Page, G.~L., Reich, B.~J., Ventrucci, M. \& Yang, S.} (2023).
			Spectral adjustment for spatial confounding.
			\textit{Biometrika} \textbf{110}, 699--719.
			
			\bibitem[Guillaumin et~al.(2022)]{Guillaumin2022}
			\textsc{Guillaumin, A.~P., Sykulski, A.~M., Olhede, S.~C. \& Simons, F.~J.} (2022).
			The debiased spatial Whittle likelihood.
			\textit{J. R. Statist. Soc. B} \textbf{84}, 1526--1557.
			
			\bibitem[Imbens \& Manski(2004)]{Imbens2004}
			\textsc{Imbens, G.~W. \& Manski, C.~F.} (2004).
			Confidence intervals for partially identified parameters.
			\textit{Econometrica} \textbf{72}, 1845--1857.
			
			\bibitem[Imbens et~al.(2025)]{Imbens2025}
			\textsc{Imbens, G.~W., Kallus, N., Mao, X. \& Wang, Y.} (2025).
			Long-term causal inference under persistent confounding via data combination.
			\textit{J. R. Statist. Soc. B} \textbf{87}, 362--388.
			
			\bibitem[Khan \& Berrett(2026)]{Khan2026}
			\textsc{Khan, K. \& Berrett, C.} (2026).
			Re-thinking spatial confounding in spatial linear mixed models.
			\textit{Statist. Sci.} \textbf{41}, 411--432.
			
			\bibitem[Khan \& Calder(2022)]{Khan2022}
			\textsc{Khan, K. \& Calder, C.~A.} (2022).
			Restricted spatial regression methods: implications for inference.
			\textit{J. Am. Statist. Assoc.} \textbf{117}, 482--494.
			
			\bibitem[K\"{u}ndig \& Sigrist(2025)]{Kundig2025}
			\textsc{K\"{u}ndig, P. \& Sigrist, F.} (2025).
			Iterative methods for Vecchia--Laplace approximations for latent Gaussian process models.
			\textit{J. Am. Statist. Assoc.} \textbf{120}, 1267--1280.
			
			\bibitem[Li et~al.(2024)]{Li2024}
			\textsc{Li, C., Sun, S. \& Zhu, Y.} (2024).
			Fixed-domain posterior contraction rates for spatial Gaussian process model with nugget.
			\textit{J. Am. Statist. Assoc.} \textbf{119}, 1336--1347.
			
			\bibitem[Llamazares-Elias et~al.(2026)]{LlamazaresElias2026}
			\textsc{Llamazares-Elias, L., Latz, J. \& Lindgren, F.} (2026).
			A parameterization of anisotropic Gaussian fields with penalized complexity priors.
			\textit{J. Am. Statist. Assoc.}, 1--12.
			
			\bibitem[Masten \& Poirier(2020)]{Masten2020}
			\textsc{Masten, M.~A. \& Poirier, A.} (2020).
			Inference on breakdown frontiers.
			\textit{Quant. Econ.} \textbf{11}, 41--111.
			
			\bibitem[{Office for National Statistics}(2021)]{ONS2021}
			\textsc{Office for National Statistics} (2021).
			Census 2021 Middle Layer Super Output Area population estimates and centroids.
			Office for National Statistics, Open Geography Portal, \texttt{https://geoportal.statistics.gov.uk/}.
			
			\bibitem[Paciorek(2010)]{Paciorek2010}
			\textsc{Paciorek, C.~J.} (2010).
			The importance of scale for spatial-confounding bias and precision of spatial regression estimators.
			\textit{Statist. Sci.} \textbf{25}, 107--125.
			
			\bibitem[Ro\v{c}kov\'{a} \& Rousseau(2024)]{Rockova2024}
			\textsc{Ro\v{c}kov\'{a}, V. \& Rousseau, J.} (2024).
			Ideal Bayesian spatial adaptation.
			\textit{J. Am. Statist. Assoc.} \textbf{119}, 2078--2091.
			
			\bibitem[Stoye(2009)]{Stoye2009}
			\textsc{Stoye, J.} (2009).
			More on confidence intervals for partially identified parameters.
			\textit{Econometrica} \textbf{77}, 1299--1315.
			
			\bibitem[Tan(2024)]{Tan2024}
			\textsc{Tan, Z.} (2024).
			Model-assisted sensitivity analysis for treatment effects under unmeasured confounding via regularized calibrated estimation.
			\textit{J. R. Statist. Soc. B} \textbf{86}, 1339--1363.
			
			\bibitem[{UK Health Security Agency}(2024)]{UKHSA2024}
			\textsc{UK Health Security Agency} (2024).
			Public Health Profiles (Fingertips): Indicators 93260, 93275, 93114 and 93240.
			UK Health Security Agency, \texttt{https://fingertips.phe.org.uk/}.
			
			\bibitem[Wiecha et~al.(2025)]{Wiecha2025}
			\textsc{Wiecha, N., Hoppin, J.~A. \& Reich, B.~J.} (2025).
			Two-stage estimators for spatial confounding with point-referenced data.
			\textit{Biometrics} \textbf{81}, ujaf093.
			
		\end{thebibliography}
\end{document}